\documentclass[AMA,STIX1COL]{WileyNJD-v2}

\usepackage{graphicx}
\usepackage{amsmath}
\usepackage{amssymb}
\usepackage{mathtools}
\usepackage{bm}
\usepackage{cancel}
\usepackage{stackengine}
\usepackage{subfig}
\usepackage[export]{adjustbox}
\usepackage{booktabs}
\usepackage{tikz}
\usetikzlibrary{matrix,decorations.pathreplacing,calc,positioning,fit}
\usepackage{array}
\usepackage{multirow}
\usepackage{tensor}

\DeclareMathOperator{\sech}{sech}

\articletype{Research Paper}%

\received{\today}
\revised{\today}
\accepted{\today}

\begin{document}

\title{Simultaneous topology and fastener layout optimization of assemblies considering joint failure}

\author[1]{Olaf Ambrozkiewicz*}

\author[1]{Benedikt Kriegesmann}

\authormark{OLAF AMBROZKIEWICZ \textsc{et al}}

\address[1]{\orgdiv{Working group structural optimization for lightweight design}, \orgname{Hamburg University of Technology}, \orgaddress{\state{Hamburg}, \country{Germany}}}

\corres{*Olaf Ambrozkiewicz \email{olaf.ambrozkiewicz@tuhh.de}}

\abstract[Summary]{This paper provides a method for the simultaneous topology optimization of parts and their corresponding joint locations in an assembly. Therein, the joint locations are not discrete and predefined, but continuously movable. The underlying coupling equations allow for connecting dissimilar meshes and avoid the need for remeshing when joint locations change. The presented method models the force transfer at a joint location not only by using single spring elements but accounts for the size and type of the joints. When considering riveted or bolted joints, the local part geometry at the joint location consists of holes that are surrounded by material. For spot welds, the joint locations are filled with material and may be smaller than for bolts. The presented method incorporates these material and clearance zones into the simultaneously running topology optimization of the parts. Furthermore, failure of joints may be taken into account at the optimization stage, yielding assemblies connected in a fail-safe manner.}


\maketitle

\section{Introduction}

Topology optimization has made its way from academia to industrial application. One of the most prominent examples is the Flight Crew Rest Compartments (FCRC) bracket of the A350, which has been replaced by an additively manufactured part (see, e.g. \cite{emmelmann_10_2017}). Often, only a single part is considered in the optimization, and its mounting points are modeled as fixed boundary conditions. In practical application, the standalone optimization of a single part has an impact on the force distribution in the structure it is connected to. Furthermore, the number and the locations of joints are predefined and therefore not optimal. For instance, the FCRC bracket is bolted at ten locations with a regular bolt pattern, which was considered as fixed for the optimization. Making the locations of bolts part of the design variables in the optimization might not only provide a better bolt pattern but also influence the result of the topology optimization of the part itself. Therefore, for a full exploitation of topology optimization it is important to optimize the parts and their connections simultaneously, i.e. to optimize an assembly as a whole.

The problem of topology optimizing parts and the layout of fasteners at the same time was tackled e.g. by Chickermane and Gea~\cite{chickermane_design_1997} in a straight forward way: As potentially force transferring joints, a grid of spring elements, acting between the nodes of two parts, is introduced. The springs are treated by the same stiffness penalization scheme (similar to RAMP~\cite{bendsoe_topology_2004}), as the continuum elements of the individual parts. By constraining the maximum ``volume'' of the springs, only the most important springs remain in the final result. However, this method does not guarantee a layout with discrete joints. Even partial joints may be stiff enough to transfer loads sufficiently good, therefore one has no control over the number of joints remaining in the final result. Also, for higher resolution FE meshes, the spring grid gets denser and the inherent lack of control over a minimum distance between joints becomes evident: If aiming for an assembly held together by multiple separate connections, a result with just a few clustered groups of joints is obtained.

To overcome the problem of non-discrete joints, recently Thomas et al.~\cite{thomas_topology_2020} studied topology and joint layout optimization using BESO~\cite{huang_evolutionary_2010} and applied their method to periodic assemblies. Nevertheless, the problem of joint clustering remains, and requires several heuristic approaches and parameters to obtain the shown results. Again, a grid of potential joints with fixed spacing is used, which might lead to suboptimal results, since joints cannot be placed at any arbitrary position.

Alternative methods for combined topology and layout optimization directly use components' location variables as part of the design vector, thus allowing flexible placement. Zhu and Zhang~\cite{zhu_integrated_2009} applied this technique for a layout optimization of fixed-shape components embedded in a topology optimized support structure. The method was later refined by including predefined, fixed fastener layouts to model the force transfer between the components and the support structure~\cite{zhu_multi-point_2015}. The fasteners are linked to the support structure using coupling equations, circumventing the need for remeshing when components move. Most recently, Rakotondrainibe et al.~\cite{rakotondrainibe_topology_2020} modeled joints as rigid supports by using movable boundary condition zones, whose positions were part of the optimization in a level-set topology optimization framework.

This paper presents a method for the density-based topology optimization of components and the simultaneous optimization of joint positions. The joints are modeled by patterns of mesh-independent springs, allowing arbitrary small movements during the optimization. Also, the physical size of a joint is accounted for in the modeling of the force transfer. Each joint additionally comes with a movable non-design space, allowing to impose restrictions on the local geometry concerning joint mountability. Bolted connections, for example, need holes to be present in the parts to be clamped. The presented method will consider these geometric features with their exact size already at the optimization stage, thus strongly coupling the topology and joint optimization. Not accounting for that in the simulations means, that holes need to be added in a later machining step, which may severely degrade the load-bearing capacity of the part. Hence, the results of the presented method will consist of exactly the required amount of material and will have exactly the desired amount of discrete joints. The method further includes fail-safe considerations into the optimization to account for the failure of single or even multiple connections.

\begin{figure}
	\centering
	\subfloat[Connection without mounting hole]{\includegraphics[width=0.475\linewidth]{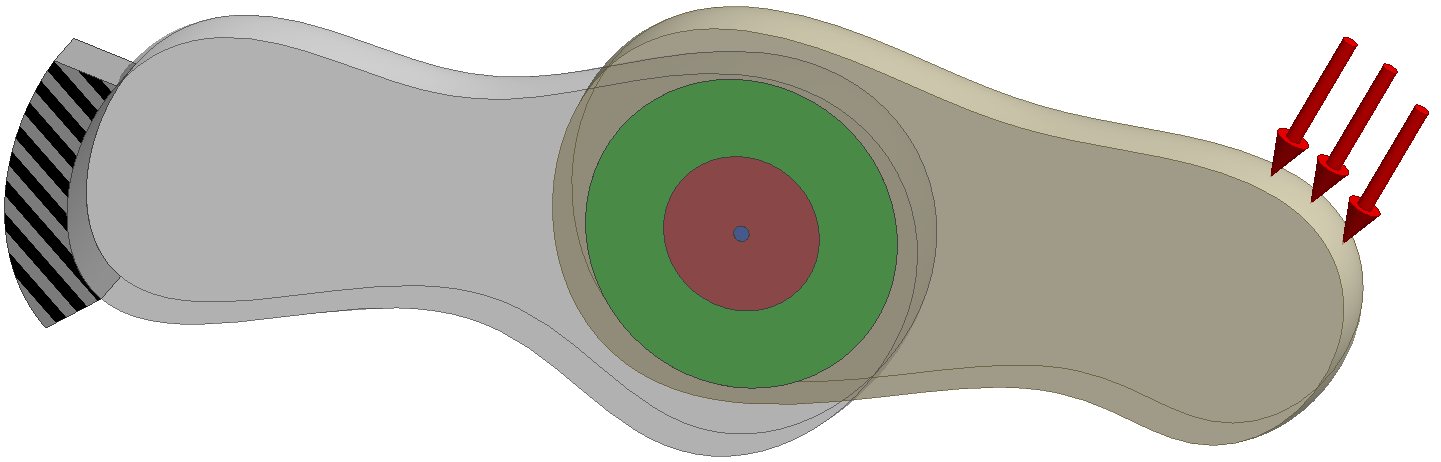}}\hfill
	\subfloat[Connection with mounting hole]{\includegraphics[width=0.475\linewidth]{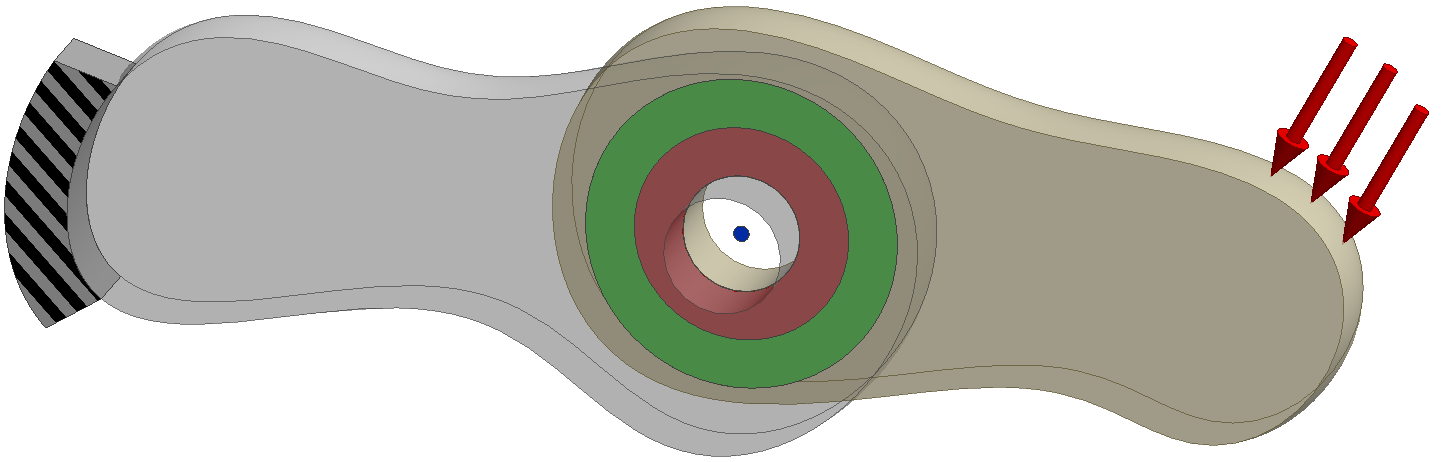}}
	\caption{Examples for a connection between two bodies. The connection is defined by a reference point (blue), a force transfer zone (red), and an additional material zone (green).}
	\label{fig:ConnTypes}
\end{figure}

Figure~\ref{fig:ConnTypes} shows the two different types of axisymmetric connections considered in this paper:

\begin{itemize}
	\item Connections that only require a minimum amount of material to be present on the mating parts
	\item Connections that additionally require mounting holes to be present in the mating parts
\end{itemize}

Examples for the first type are spot welds or the spot-wise application of adhesives. Connections using bolts or rivets are examples for the other type of connection, where holes of a specific diameter must be present in both parts, together with some amount of surrounding material for the load transfer.

Figure~\ref{fig:ConnTypes} further shows the two geometric zones, that define the connection: The force transfer zone, shown in red, in which forces act between the two bodies and the minimum material zone, which extends the force transfer zone by an additional amount of material, shown in green. Both zones lie on the contact surface of the bodies. A reference point (shown in blue) defines the location of the joint on the contact surface.

This paper outlines as follows: Section~\ref{sec:Approach} covers the presented method briefly in discussing the design variables, the optimization problem, what parts the system matrix is composed of, and how individual variables relate to each other. More details are given in the following sections: Section~\ref{sec:TO} recapitulates the equations for the SIMP approach, used for the topology optimization of the parts. For multi-component optimization, a slight modification of the governing equations is needed. Section~\ref{sec:NDS} explains, how the minimum material zones and the holes are implemented as movable non-design spaces in the topology optimization framework. The modeling of the force transfer through the joints is covered in section~\ref{sec:JOpt}. The objective and constraint functions used in the optimization problem are presented in section~\ref{sec:optimization}. Numerical examples in 2D and 3D are given in sections~\ref{sec:ex2D} and \ref{sec:ex3D}. A summary and outlook are given in section~\ref{sec:outlook}.

\section{Overall approach}\label{sec:Approach}

\begin{figure}
	\centering
	\subfloat[Numerical model\label{fig:Example_DS2D}]{\includegraphics[trim={30, 50, 10, 40},clip,width=0.475\linewidth]{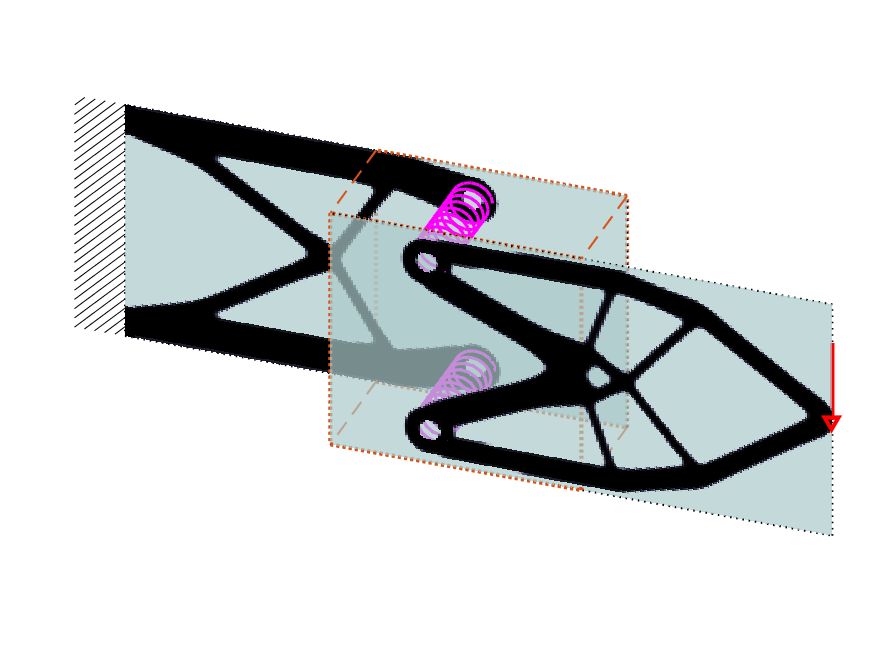}}\hfill
	\subfloat[Physical model with fasteners]{\includegraphics[trim={20, 25, 80, 30},clip,width=0.475\linewidth]{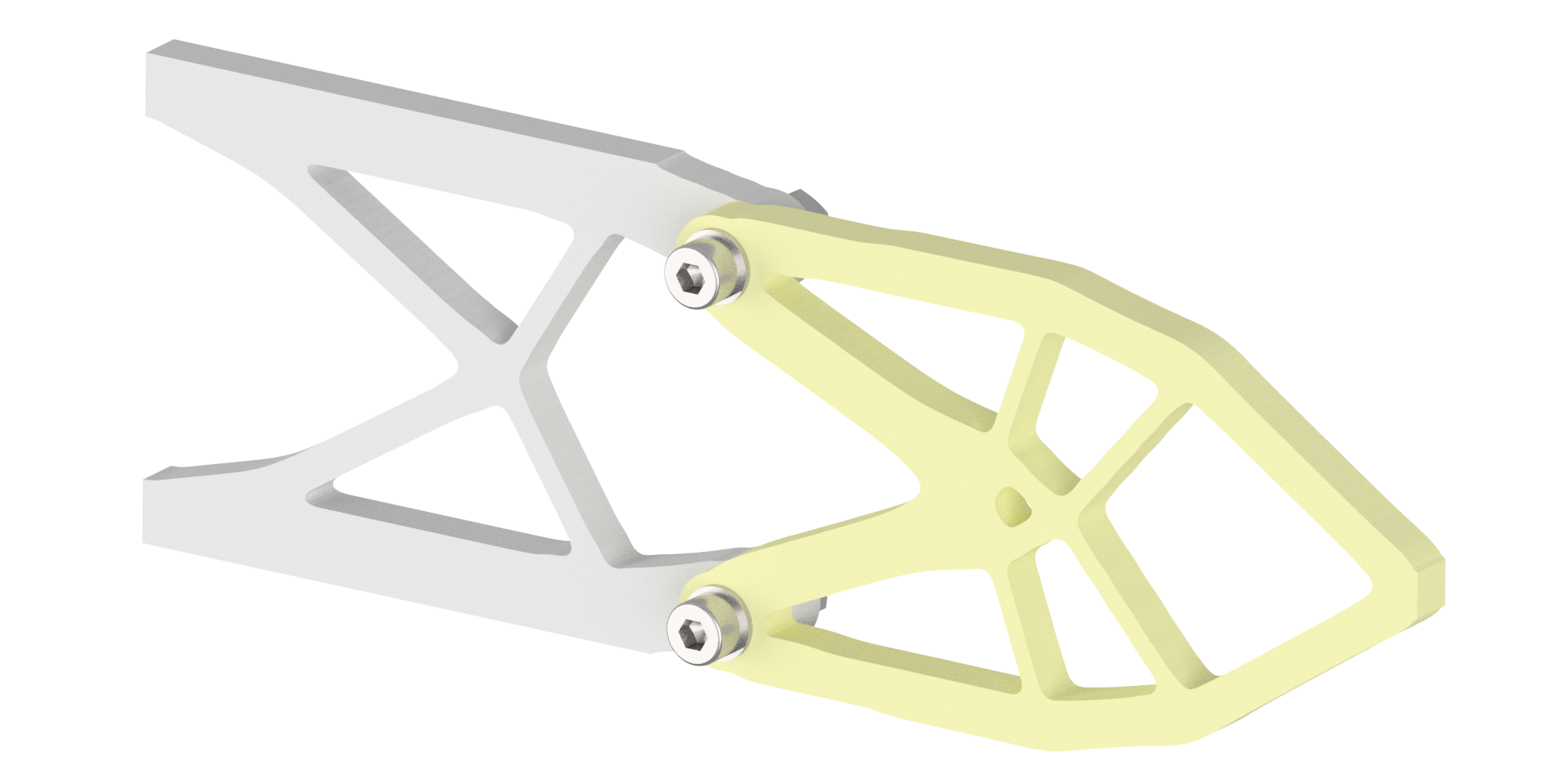}}
	\caption{Result for simultaneous topology and joint optimization.}
	\label{fig:Example_2D}
\end{figure}

The design space of each of the $n_p$ parts is discretized by the finite element method. The SIMP approach~\cite{bendsoe_optimal_1989} is used to simultaneously optimize the material distribution for every part. Each part has an own, independent design space, shown as layers in figure~\ref{fig:Example_DS2D}. No nodes are shared between the parts' meshes.

The design variables of the $n_j$ joints are the locations of the reference points on the contact surface (orange zones in figure~\ref{fig:Example_DS2D}). A solid material zone belongs to every joint (cf. figure~\ref{fig:ConnTypes}). Therefore, circular non-design spaces are included in the material distribution field of each connected part at the location of the joints.

The fasteners, connecting the parts, are simplified to patterns of springs, modeling the equivalent stiffness and the force transfer area's size of the real fastener (cf. figure~\ref{fig:Example_2D}). The properties and the number of fasteners are predefined and do not change within the optimization. However, the springs move according to the corresponding joint's reference point and have their own independent nodes. Coupling equations link the nodes of the connection springs to the nodes of the parts, allowing force transfer between the parts.

\subsection{Design variables}\label{sec:DVs}

The design variables for the topology optimization are concatenated in a vector $\bm{\varrho}$ with one entry for every of the $n_e$ elements of the total FE model:

\begin{equation}
\bm{\varrho} = \begin{bmatrix}
\varrho_1\\ \vdots \\ \varrho_{n_e}
\end{bmatrix}
\end{equation}

The design variables of the joints are the locations of their reference points. Without loss of generality, a plane contact surface parallel to the global $x$-$y$-plane is considered in this paper. The position vector $\bm{x}^i$ of each of the $n_j$ joints is therefore a vector with two entries (no rotation is considered here):

\begin{equation}
\bm{x}^i = \begin{bmatrix}
x^i \\ y^i\end{bmatrix},\qquad i = 1, \dots, n_j
\end{equation}

The design vector $\bm{x}$ contains all of the joints' positions:

\begin{equation}
\bm{x} = \begin{bmatrix}
\bm{x}^1 \\ \vdots \\  \bm{x}^{n_j} \end{bmatrix}
\end{equation}

\subsection{Optimization problem}

The investigated optimization problem, with an objective function $c$ and inequality constraints $h_i$, states:

\begin{subequations}
	\begin{align}
	&\displaystyle{\min_{\bm{\varrho},\bm{x}} c\left(\tilde{\bm{u}}(\bm{\varrho},\bm{x})\right)} &&\\
	\rm{s.t.}\qquad &h_i \leq 0, &&\forall i\\
	&0 \leq \varrho_j \leq 1,  &&j = 1, \dots, n_e\label{eq:rhobounds}\\
	&\!\begin{rcases}
	x_l \leq x^k \leq x_u\\
	y_l \leq y^k \leq y_u
	\end{rcases},  &&k = 1,\dots,n_j\label{eq:xbounds}\\
	&\tilde{\bm{K}} \tilde{\bm{u}} = \tilde{\bm{f}}\label{eq:Kuf}&&
	\end{align}
\end{subequations}

The objective function $c$ is dependent on the augmented displacement vector $\tilde{\bm{u}}$ that in turn is the solution to the equilibrium equation~(\ref{eq:Kuf}) of linear elastic bodies, discretized by FEM. The optimization is further constrained by one or more constraint functions $h_i$. Objective and constraint functions, as well as their derivatives, are discussed in section~\ref{sec:optimization}.

Equations (\ref{eq:rhobounds}) and (\ref{eq:xbounds}) employ bounds on the values of the design variables. The material design variables $\bm{\varrho}$ are limited to a value between 0 and 1. The position variables in $\bm{x}$ are constrained with upper and lower bounds, such that the joints, including their material zones, stay inside the designated contact surface area of the parts.

The optimization problem is solved using the method of moving asymptotes (MMA)~\cite{svanberg_method_1987}.

\subsection{System matrix}\label{ss:system}

The augmented stiffness matrix $\tilde{\bm{K}}$ from equation~(\ref{eq:Kuf}) is composed of the stiffness contribution $\bm{K}_m$, originating from the material of the parts, the stiffness $\bm{K}_c$, modeling the connections, and linear coupling equations, grouped in $\bm{G}$:

\begin{subequations}
	\begin{align}
	\tilde{\bm{K}} \tilde{\bm{u}} &= \tilde{\bm{f}}\\
	\left(\bm{K}_m(\bm{\varrho}, \bm{x}) + \bm{K}_c + \bm{G}(\bm{x})\right) \tilde{\bm{u}} &= \tilde{\bm{f}}\label{eq:KmKcG}
	\end{align}
\end{subequations}

Since every joint comes with a material zone as non-design space, the material part $\bm{K}_m$ is not only dependent on the material design variables $\bm{\varrho}$ but also on the joint locations, stored in $\bm{x}$. The matrix $\bm{K}_c$ of the joints is constant since the properties of each connection are predefined. The coupling equations in $\bm{G}$ control, where the individual joints are placed on the parts and are therefore solely dependent on the $\bm{x}$ design vector.

All parts and all joints are independent of each other and do not share degrees of freedom. The individual stiffness matrices $\bm{K}_m^i$ of each part $i$ are therefore placed on the diagonal of $\tilde{\bm{K}}$. The same accounts for the stiffness matrices $\bm{K}_c^j$ of each joint $j$. The augmented stiffness matrix $\tilde{\bm{K}}$ has therefore the shape shown in equation~(\ref{eq:Kmatrix}). Only the linear coupling equations in $\bm{G}$ link the degrees of freedom of the joints and the parts, such that a force transfer between the parts is possible.

\begin{equation}\tilde{\bm{K}} = 
{\def\arraystretch{1.2}
	\begin{bmatrix}
	\begin{array}{cc}
	\begin{array}{cc}
	\begin{bmatrix}
	\bm{K}_m^1 & & \bm{0}\\
	& \ddots & \\
	\bm{0} & & \bm{K}_m^{n_p}
	\end{bmatrix} 	& \bm{0}\\ 
	\bm{0} & \begin{bmatrix}
	\bm{K}_c^1 & & \bm{0}\\ 
	& \ddots & \\
	\bm{0} & & \bm{K}_c^{n_j}
	\end{bmatrix}  
	\end{array} & {\def\arraystretch{1.4}\begin{bmatrix}
		\\
		\\
		\bm{G}^T\\
		\\
		\\
		\end{bmatrix}} \\ 
	\boldsymbol[{\addtolength{\arraycolsep}{21.0pt}
		\begin{matrix}
		\phantom{0} & \bm{G}& \phantom{0}\\
		\end{matrix}}\boldsymbol]& \bm{0}
	\end{array}
	\end{bmatrix}
}\label{eq:Kmatrix}
\end{equation}

The augmented displacement vector $\tilde{\bm{u}}$ contains the nodal displacements of the parts $\bm{u}_m$ and the joints $\bm{u}_c$, as well as the Lagrange-multipliers $\bm{\lambda}_c$ for the coupling equations:

	\begin{equation}
	\tilde{\bm{u}} = \begin{bmatrix}
	\bm{u}_m \\ \bm{u}_c \\ \bm{\lambda}_c
	\end{bmatrix}, \textrm{with:}\qquad
	\bm{u}_m = \begin{bmatrix}
	\bm{u}_m^1 \\ \vdots \\ \bm{u}_m^{n_p}
	\end{bmatrix}, \quad
	\bm{u}_c = \begin{bmatrix}
	\bm{u}_c^1 \\ \vdots \\ \bm{u}_c^{n_j}
	\end{bmatrix}, \quad
	\bm{\lambda}_c = \begin{bmatrix}
	\bm{\lambda}_c^1 \\ \vdots \\ \bm{\lambda}_c^{n_j}
	\end{bmatrix}
	\end{equation}

There are as many Lagrange multipliers in $\bm{\lambda}_c$ as entries in $\bm{u}_c$, since all degrees of freedom of the joints are coupled by an own coupling equation.

Assuming, that external loads only act on the nodes of the parts, the augmented force vector $\tilde{\bm{f}}$ is:

\begin{equation}
\tilde{\bm{f}} = \begin{bmatrix}
\bm{f}_m\\ \bm{0} \\ \bm{0}
\end{bmatrix}, \textrm{with:}\qquad\bm{f}_m = \begin{bmatrix}
\bm{f}_m^1 \\ \vdots \\ \bm{f}_m^{n_p}
\end{bmatrix}
\end{equation}


\subsection{Relation of variables}

\begin{figure}
	\centering
	\begin{tikzpicture}
	\node[draw, align=center, rounded corners=4pt](DVm) at (0,0) {Material DV\\$\bm{\varrho}$};
	\node[draw, align=center, rounded corners=4pt](DVx) at (5.5,0) {Joint DV\\$\bm{x}$};
	\node[draw, align=center, rounded corners=4pt](M) at (4,-1.2) {Masks for NDS\\$\bm{\psi}(\bm{x})$};
	\node[draw, align=center, rounded corners=4pt](rp) at (1,-1.2) {Filtering, projection\\$\tilde{\bm{\varrho}}$, $\bar{\bm{\varrho}}$};
	\node[draw, align=center, rounded corners=4pt](rm) at (1.5,-2.4) {Modified densities\\$\hat{\bm{\varrho}}(\bm{\varrho},\bm{x})$};
	\node[draw, align=center, rounded corners=4pt](Km) at (1.5,-3.6) {Material\\$\bm{K}_m(\bm{\varrho},\bm{x})$};
	\node[draw, align=center, rounded corners=4pt](G) at (5.5,-3.6) {Coupling\\$\bm{G}(\bm{x})$};
	\node[draw, align=center, rounded corners=4pt](Kc) at (3.5,-3.6) {Joint\\$\bm{K}_c$};
	\node[draw, align=center, rounded corners=4pt](Ka) at (3.5,-4.8) {System matrix\\$\tilde{\bm{K}}(\bm{\varrho},\bm{x})$};
	
	\draw[->] (DVm) to (rp);
	\draw[->] (DVx) to (M);
	\draw[->] (DVx) to (G);
	\draw[->] (rp) to (rm);
	\draw[->] (M) to (rm);
	\draw[->] (rm) to (Km);
	\draw[->] (Km) to (Ka);
	\draw[->] (G) to (Ka);
	\draw[->] (Kc) to (Ka);
	\end{tikzpicture}
	\caption{Relation of the variables.}
	\label{fig:Relation}
\end{figure}
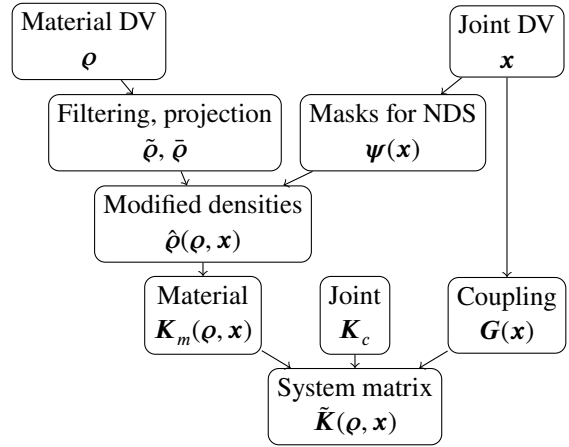

Figure~\ref{fig:Relation} shows, how the different variables are related to each other.

On the topology optimization side, the material design variables (DV) $\bm{\varrho}$ yield filtered and projected variables, which is covered in section~\ref{sec:TO}. The joint position variables $\bm{x}$ are needed to define mask vectors $\bm{\psi}$ that are used to employ the material zones of each joint as a non-design space (NDS), as explained in section~\ref{sec:NDS}. A field of ``modified densities'' combines the material distribution obtained by topology optimization with the NDS of the joints and is the input for the material part $\bm{K}_m$ of the global system matrix.

On the joint optimization side, section~\ref{sec:JOpt} shows, how the joints' position variables $\bm{x}$ influence the coupling matrix $\bm{G}$, that links the connection springs to the parts. The joints itself are modeled by patterns of springs with a constant predefined stiffness. All joints are combined to the connection part $\bm{K}_c$ of the global stiffness matrix.

\section{Fields used for topology optimization}\label{sec:TO}

When using the SIMP approach with projection, three fields of material variables are present~\cite{sigmund_topology_2013}: The design variables $\bm{\varrho}$, the filtered variables $\tilde{\bm{\varrho}}$ (see section~\ref{ss:filtered}), and the projected variables $\bar{\bm{\varrho}}$ (see section~\ref{ss:projected}). 

In this paper, also a fourth field of ``modified densities'' $\hat{\bm{\varrho}}$ is needed, employing the non-design space material zones of the joints in the model (see section~\ref{ss:modified}).

\subsection{Filtered variables}\label{ss:filtered}

A filtering step is applied to the material design variables $\bm{\varrho}$, which yields the filtered variables $\tilde{\bm{\varrho}}$~\cite{bourdin_filters_2001, bruns_topology_2001}.

For multi-component topology optimization the filtering has to be applied on a per-part basis, otherwise, material would be transferred between different parts in areas, where the FE meshes overlap or contact each other, hindering the parts to evolve independently. Therefore, the meshes of every part are treated as independent ``layers'' and the filter equation is only evaluated over a set $\mathbb{L}_k$ containing the element indices of the current part $k$:

\begin{subequations}
	\begin{align}
	\tilde{\varrho}_i &= \frac{\sum_{j \in \mathbb{L}_k} w(\bm{c}^j - \bm{c}^i)v_j \varrho_j}{\sum_{j \in \mathbb{L}_k} w(\bm{c}^j - \bm{c}^i)v_j},\qquad k = 1,\dots,n_p\label{eq:filter}\\
	w(\bm{x}') &\coloneqq \max(r-\lVert \bm{x}' \rVert_2, 0)
	\end{align}
\end{subequations}

In equation~(\ref{eq:filter}), $v_j$ and $\bm{c}^j$ refer to the volume and the center point of an element $j$, $w$ is a conic weighting function dependent on the the filter radius $r$.

\subsection{Projected variables}\label{ss:projected}

The third field of material variables are the projected variables $\bar{\bm{\varrho}}$ that are obtained by applying a differentiable approximation of the Heaviside step-function to the filtered variables~\cite{wang_projection_2011}:

\begin{equation}
\bar{\varrho}_i = \frac{\tanh(\beta \eta)+\tanh(\beta (\tilde{\varrho}_i- \eta))}{\tanh(\beta \eta)+\tanh(\beta(1-\eta))}\label{eq:projection}
\end{equation}

The steepness parameter $\beta$ is gradually increased during the optimization, the threshold parameter $\eta$ is set to a constant value of $\eta=0.5$.

\subsection{Modified densities}\label{ss:modified}

The fourth field are the modified densities $\hat{\bm{\varrho}}$ that are dependent on the projected variables and the locations of the joints:

\begin{equation}
\hat{\bm{\varrho}} = \hat{\bm{\varrho}}\left(\bar{\bm{\varrho}},\bm{x}\right)
\end{equation}

The modified densities employ the moving non-design spaces, originating from the material and clearance zones of the joints, into the material distribution field obtained by topology optimization (for details see section~\ref{sec:NDS}), thus creating a link between the material and the position design variables.

The modified densities $\hat{\bm{\varrho}}$ are the physical densities used to interpolate the effective Young's modulus $E_i$ of each element $i$ between a value $E_0$ for the solid and $E_{min} \ll E_0$ for the void material phase, according to the SIMP scheme~\cite{bendsoe_optimal_1989}:

\begin{equation}
E_i = E_{min}+(E_0-E_{min})(\hat{\varrho}_i)^p,\qquad\hat{\varrho}_i\in [0,1]
\end{equation}

The exponent $p > 1$ is used to penalize intermediate densities.

\section{Moving non-design spaces}\label{sec:NDS}

The non-design spaces (NDS) are used to employ solid material zones or holes of specific sizes for every joint. In this paper, only circular (or cylindrical) features are considered (cf. figure~\ref{fig:ConnTypes}). The NDS follow the joints' reference points and therefore need to be continuously movable, independently of the discretization of the underlying FE mesh.

The NDS are defined by parametric shapes that are mapped to a general mask vector $\bm{\psi}$~\cite{ambrozkiewicz_density-based_2020}.
The masks can either be used to remove material (denoted by $\tensor[^-]{\bm{\psi}}{}$), add material (denoted by $\tensor[^+]{\bm{\psi}}{}$), or do both by subsequent application. These three cases are discussed in sections~\ref{ss:voidNDS} --~\ref{ss:ringNDS}.

A total mask $\bm{\psi}$ is composed of individual masks $\bm{\psi}^i$, obtained for every joint $i$. Section~\ref{ss:singleMask} describes, how a single mask is calculated, section~\ref{ss:maskCombo} covers the combination of masks.

\subsection{Mask for a single axisymmetric feature}\label{ss:singleMask}

The mask vector $\bm{\psi}^i$ for joint $i$ has one entry for every of the $n_e$ finite elements of the model and has, therefore, the same size as the material design vector $\bm{\varrho}$.

The joint's mask will define a circular shape (or cylindrical shape in 3D) around the joint's reference point in the $x$-$y$-plane, which is considered as the contact plane here. The entries of $\bm{\psi}^i$ have values between 0 and 1 and are calculated for every element $j$~\cite{ambrozkiewicz_density-based_2020}:

\begin{equation}
\psi^i_j = \frac{\tanh{\left(\alpha E(\bm{c}^j - \bm{x}^i)\right)}+1}{2}\label{eq:maskval}
\end{equation}

Where $\bm{c}^j$ is the position vector of the element's center point and $E$ is the function describing the desired shape with the outline defined at $E=0$.

Equation~(\ref{eq:maskval}) maps element locations inside the shape ($E<0$) to a value of 0, and locations outside ($E>0$) to a value of 1, in a differentiable way. The parameter $\alpha$ controls the sharpness of the transition, $\alpha = 10$ is used throughout the paper.

The circular shapes are described by the function:

\begin{equation}
E(\bm{x}') := \left(\frac{x'}{r'}\right)^2 + \left(\frac{y'}{r'}\right)^2 -1\label{eq:circ}
\end{equation}

Depending on the case, for which geometric feature the mask is used, $r'$ is either the radius of the solid material zone or the radius of a hole.

Since equation~(\ref{eq:maskval}) is differentiable with respect to the joint's reference point coordinates $\bm{x}^i$, the derivatives of the mask can be included in a gradient-based optimization. The derivatives are given in Appendix~\ref{ap:maskderiv}.

\subsection{Combined masks and complementary masks}\label{ss:maskCombo}

The masks $\bm{\psi}^i$, obtained for every joint $i$, can be composed into a resulting mask  $\bm{\psi}$ by element-wise multiplication:

\begin{equation}
\bm{\psi} = \prod_i{\bm{\psi}^i},\qquad i = 1,\dots,n_j\label{eq:maskcombined}
\end{equation}

The derivatives of the combined mask with respect to the coordinates of the reference point of the $i$-th joint are:

\begin{subequations}
	\begin{align}
	\frac{\partial \bm{\psi}}{\partial x^i} &= \frac{\partial \bm{\psi}^i}{\partial x^i} \prod_{j\neq i}{\bm{\psi}^j}\\
	\frac{\partial \bm{\psi}}{\partial y^i} &= \frac{\partial \bm{\psi}^i}{\partial y^i} \prod_{j\neq i}{\bm{\psi}^j}
	\end{align}
\end{subequations}

The complementary mask is obtained by:

\begin{equation}
\bm{\psi}_{comp} = \bm{I}-\bm{\psi}
\end{equation}

\begin{figure}
	\centering
	\subfloat[Mask $\bm{\psi}$\label{fig:MaskHole}]{\includegraphics[width=0.2\linewidth]{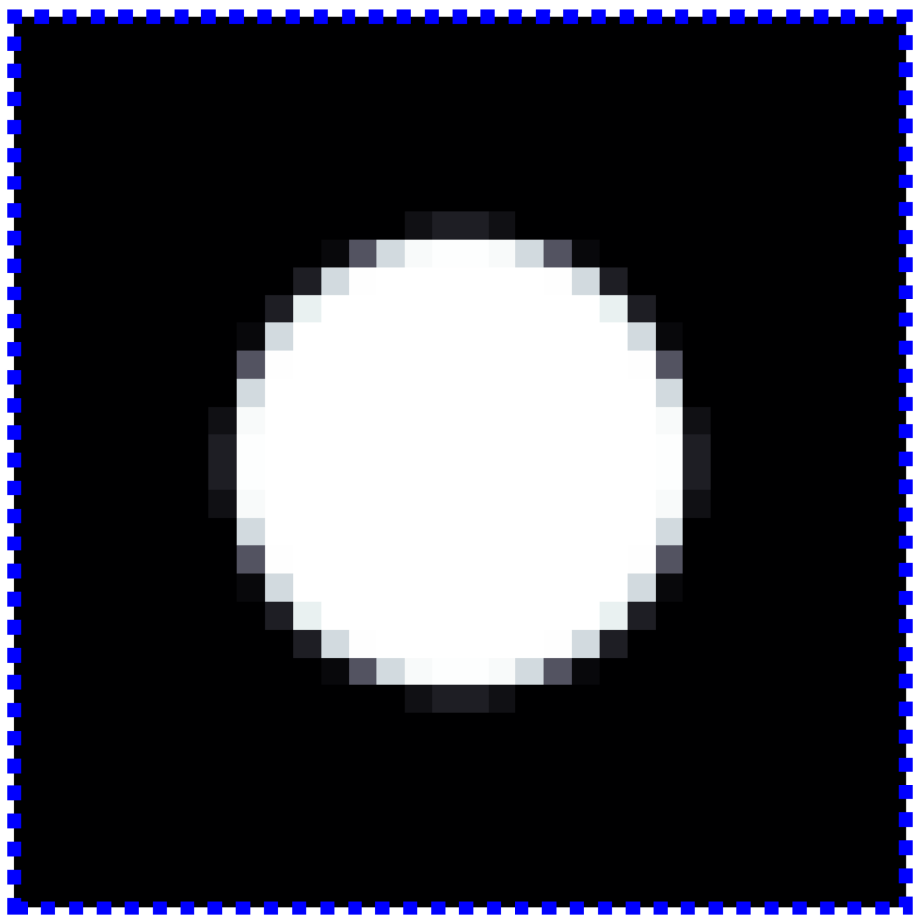}}\qquad
	\subfloat[Complementary mask $\bm{\psi}_{comp}$\label{fig:MaskCirc}]{\includegraphics[width=0.2\linewidth]{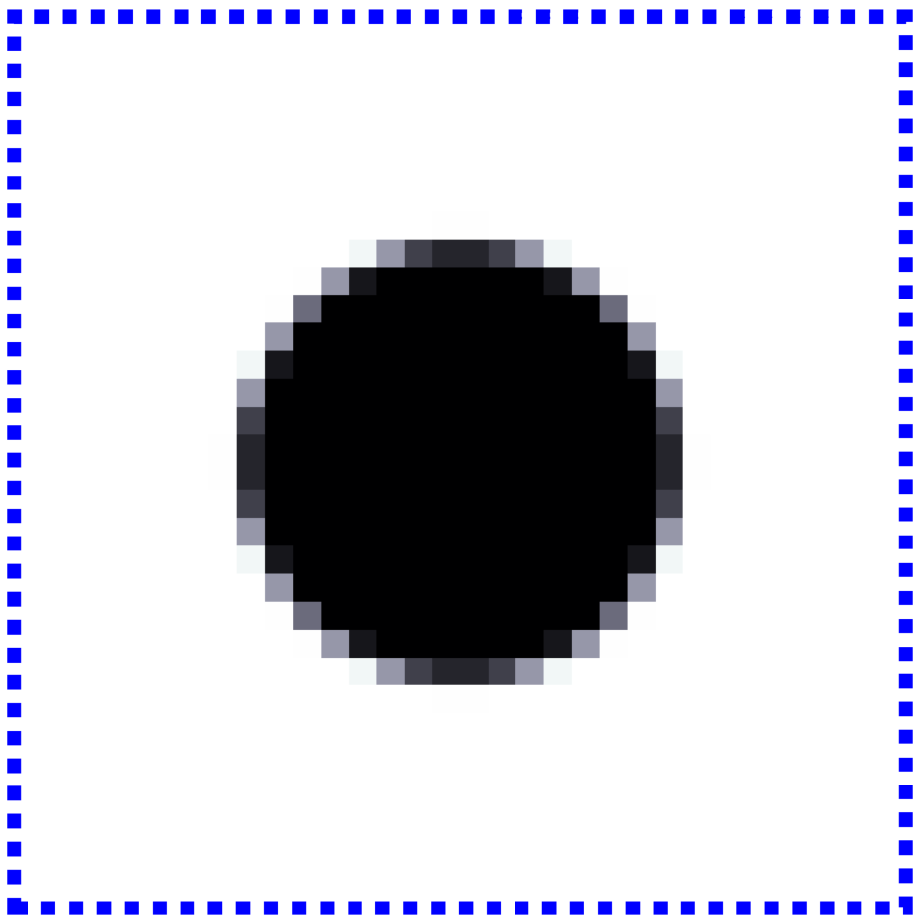}}
	\caption{A mask and its complementary mask.}
	\label{fig:Masks}
\end{figure}

Figure~\ref{fig:Masks} shows a mask and its complementary mask for a single circle with a radius of 8 inside a $32 \times 32$ domain. The parameter $\alpha$ is set to a value of 10.

\subsection{Employing void NDS}\label{ss:voidNDS}

In practical applications, holes have to be present to mount fasteners like bolts or rivets. Following the method of the authors~\cite{ambrozkiewicz_density-based_2020}, a multiplicative mask is used to employ the holes in the structure. The holes move according to the reference points and always enforce that the affected regions are kept free of solid material.

With the combined mask vector defined in equation~(\ref{eq:maskcombined}), several holes can be introduced at once by element-wise multiplication:

\begin{equation}
\bm{\hat{\varrho}} = \bm{\bar{\varrho}} \circ \tensor[^-]{\bm{\psi}}{}\label{eq:maskhole}
\end{equation}

The values of the mask $\tensor[^-]{\bm{\psi}}{}$ linearly interpolate the modified density values between zero and the value of the original projected variables.

The joint's mask vectors are obtained by applying equations~(\ref{eq:maskval}) and (\ref{eq:circ}) with the desired hole radius. The sensitivities with respect to the projected densities and the joints' locations are given in Appendix~\ref{ap:holederiv}.

The application of the mask from figure~\ref{fig:MaskHole} to enforce a hole according to equation~(\ref{eq:maskhole}) in an exemplary gradual density field is shown on the left side of figure~\ref{fig:MaskEffect}.

\subsection{Employing solid NDS}\label{ss:solidNDS}

Let $\bm{\bar{\varrho}}_{comp}$ and $\bm{\hat{\varrho}}_{comp}$ denote the complementary field of the projected variables and modified densities, where solid and void regions are interchanged:

\begin{equation}
\bm{\bar{\varrho}}_{comp} = \bm{I} - \bm{\bar{\varrho}},\qquad\bm{\hat{\varrho}}_{comp} = \bm{I} - \bm{\hat{\varrho}}\label{eq:comprho}
\end{equation}

The opposite effect of enforcing voids is achieved, if the method from section~\ref{ss:voidNDS} is applied to the complementary projected variable field:

\begin{equation}
\bm{\hat{\varrho}}_{comp} = \bm{\bar{\varrho}}_{comp} \circ \tensor[^+]{\bm{\psi}}{}\label{eq:maskcirccomp}
\end{equation}

If the result is then inverted again, a material zone is added to the density field. Therefore, the mask is now referred to as $\tensor[^+]{\bm{\psi}}{}$, since its purpose is to bring in material. Rearranging equation~(\ref{eq:maskcirccomp}) by using (\ref{eq:comprho}) yields:

\begin{subequations}
	\begin{align}
	\bm{\hat{\varrho}} &= \bm{I} - \bm{\bar{\varrho}}_{comp} \circ \tensor[^+]{\bm{\psi}}{}\\
	&= \bm{I} - (\bm{I}-\bm{\bar{\varrho}}) \circ \tensor[^+]{\bm{\psi}}{}\\
	&= (\bm{I}-\tensor[^+]{\bm{\psi}}{}) + \bm{\bar{\varrho}} \circ \tensor[^+]{\bm{\psi}}{}\label{eq:maskcirc2}\\
	&= \tensor[^+]{\bm{\psi}}{_{comp}} + \bm{\bar{\varrho}} \circ \tensor[^+]{\bm{\psi}}{}\label{eq:maskcirc}
	\end{align}
\end{subequations}

Equation (\ref{eq:maskcirc2}) shows, that the components of the mask $\tensor[^+]{\bm{\psi}}{}$ linearly interpolate the modified densities between the value one and the original values in $\bm{\bar{\varrho}}$. The derivatives are given in Appendix~\ref{ap:circderiv}.

The joints mask values are again obtained by equations~(\ref{eq:maskval}) and (\ref{eq:circ}), but in this case, the desired solid material zone radius is used.

The application of the masks from figure~\ref{fig:Masks}, according to equation~(\ref{eq:maskcirc}), adds a material circle to the gradual sample field in the middle of figure~\ref{fig:MaskEffect}.

\subsection{Employing ring-shaped solid NDS}\label{ss:ringNDS}

Subsequent application of the procedures from section~\ref{ss:solidNDS} and \ref{ss:voidNDS} yields material rings (or hollow cylinders in 3D): First a circular material zone with an larger outer radius is added and then the material inside an inner radius zone is removed:

\begin{equation}
\bm{\hat{\varrho}} = (\tensor[^+]{\bm{\psi}}{_{comp}} + \bm{\bar{\varrho}} \circ \tensor[^+]{\bm{\psi}}{}) \circ  \tensor[^-]{\bm{\psi}}{}\label{eq:maskring}
\end{equation}

\begin{figure}
	\centering
	\includegraphics[width=0.5\linewidth]{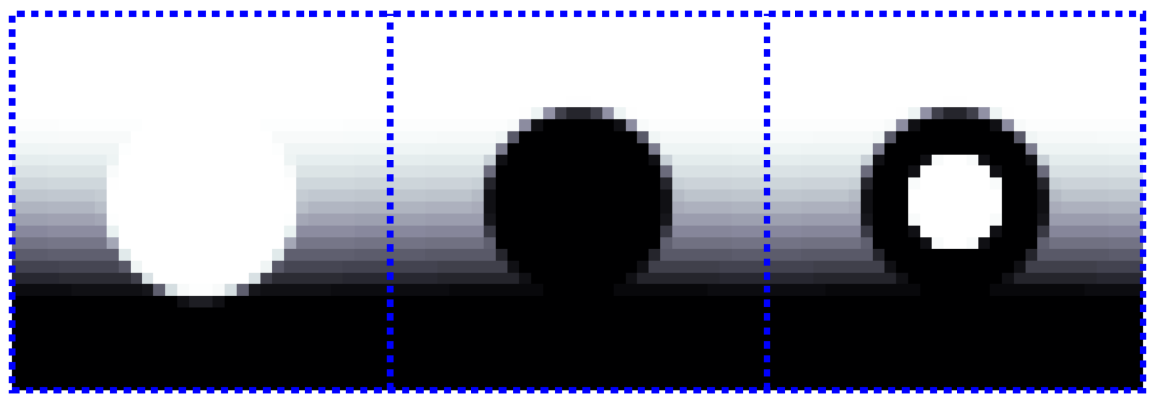}
	\caption{Application of masks to produce a hole, a circle and a ring in a sample density field.}
	\label{fig:MaskEffect}
\end{figure}

The derivatives of the above equation are given in Appendix~\ref{ap:ringderiv}. Using the masks from figure~\ref{fig:Masks} to add material, and a mask with half the radius to remove material from the sample density field, yields the result shown on the right side of figure~\ref{fig:MaskEffect}.

\section{Force transfer in connections}\label{sec:JOpt}

\begin{figure}
	\centering
	\includegraphics[width=0.35\linewidth]{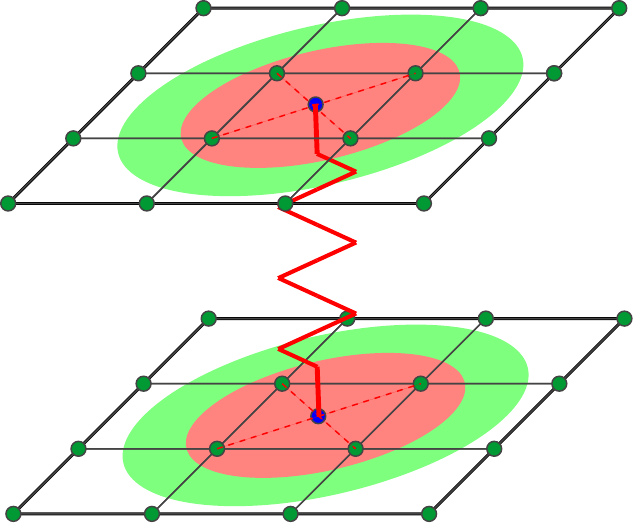}
	\caption{Mesh independent spring (in all DOF) with dependent nodes (blue) coupled to the nodes of the parts' meshes.}
	\label{fig:Layers}
\end{figure}

The force transfer between the connected parts is done by inserting stiff springs, located inside the force transfer zone of every joint, as shown in figure~\ref{fig:Layers}. Just like the NDS discussed in section~\ref{sec:NDS}, these springs also follow the joints' reference points and need to be continuously movable across the FE mesh.

\subsection{Mesh independent springs}

For every spring a new dependent ``slave'' node is inserted at the exact location of the spring on the contact surface of every part (see figure~\ref{fig:Layers}). The dependent nodes are coupled to the nodes of the corresponding element, which act as ``master'' nodes, to ensure displacement compatibility. Using this method, the meshes of the connected parts do not need to be congruent at the contact surface.

The inserted springs have the same stiffness in all degrees of freedom to model isotropic force transfer. The implementation of orthotropic behavior with e.g. different normal than tangential stiffness is straight forward.

\subsection{Spring patterns}

Since using only one spring per joint lacks the modeling of the physical size of the area where a joint can transfer load, distributed patterns of springs are used for every connection. The whole pattern will move as a unit, according to the movement of the joint's reference point. 

\begin{figure}
	\centering
	\subfloat[Circular pattern\label{fig:CircPattern2D}]{\includegraphics[width=0.2\linewidth]{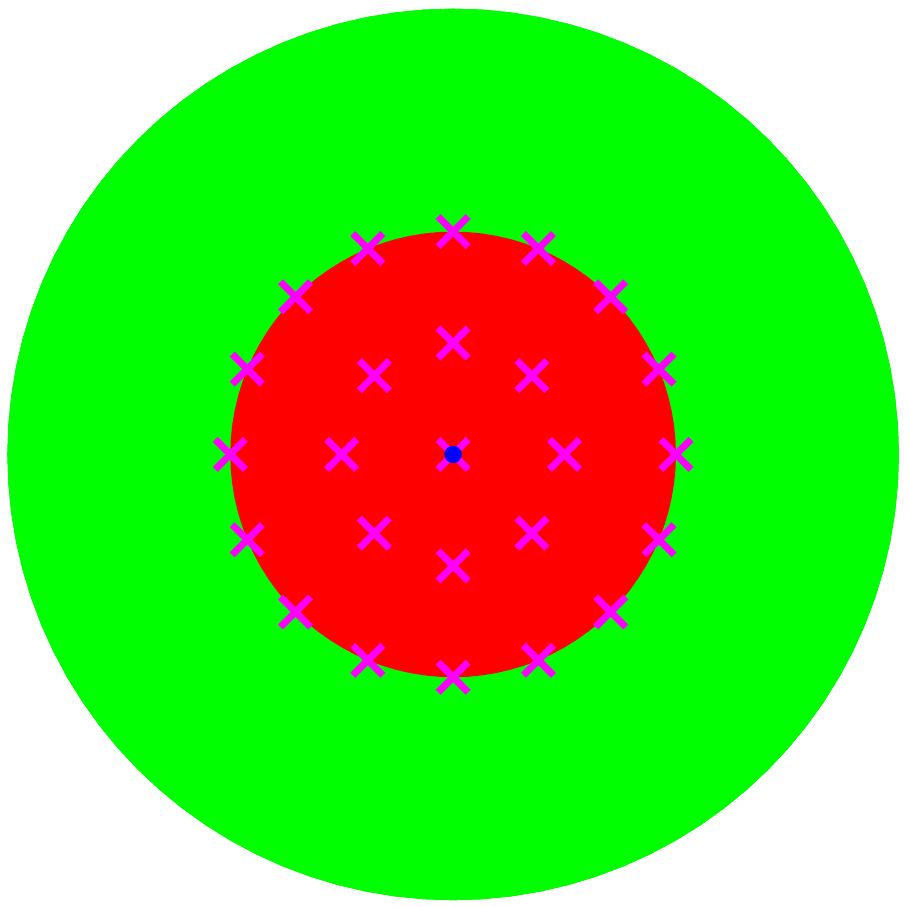}}\qquad
	\subfloat[Ring pattern\label{fig:RingPattern2D}]{\includegraphics[width=0.2\linewidth]{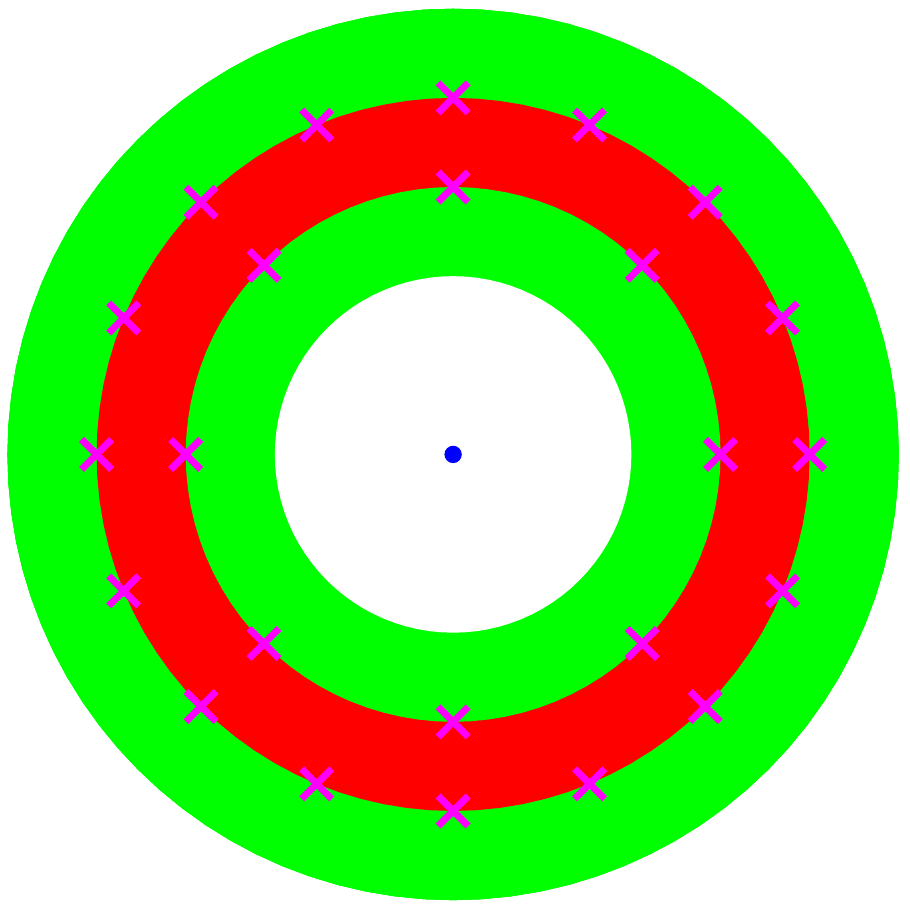}}
	\caption{NDS zone (green), force transfer zone (red), reference points (blue), and spring patterns (magenta) used in the examples.}
	\label{fig:Patterns2D}
\end{figure}

For the examples considered in this paper, concentric patterns of springs are used. Figure~\ref{fig:CircPattern2D} shows the model used for connections that do not require a hole, like e.g. spot welds. As NDS, a solid material zone (green) is present for this type of connection. A subarea of the material zone is the force transfer zone (red), over which 25 springs (including one at the central reference point) are distributed. For connections requiring a hole, the pattern from figure~\ref{fig:RingPattern2D}, consisting of 24 springs, is used.

For both cases, the resultant stiffness of the spring patterns has the same value $k_c$. The stiffness $k_c$ is chosen such, that it models the stiffness of the real connection element. In this paper, it is assumed, that the total compliance of the assembly is dominated by the compliance of the individual parts and that the connectors are comparably stiff. Therefore, stiff springs are used.

\subsection{Coupling equations}

The coupling of the dependent spring nodes to the master nodes of the parts (see figure~\ref{fig:Layers}) is described via linear equations, linking the displacements of the nodes.

Compatibility of the displacements in ensured, if the dependent node of the joint's spring is linked to the corresponding element's nodal displacements via the isoparametric weight coefficients:

\begin{equation}
\bm{g}^i = \sum_j w_j \bm{u}^j_{m} - \bm{u}^i_c = \bm{0}\label{eq:MPC}
\end{equation}

Here, $\bm{u}^i_c$ is a slave node's displacement vector belonging to a connector spring of the $i$-th joint, and $\bm{u}^j_{m}$ are the corresponding element's $j$-th nodal displacement vectors.

The weights $w_j$ in equation~(\ref{eq:MPC}) are the values of the $n_n$ element's shape functions at the position of the slave node. The values are obtained by transforming the global coordinates of the slave node to the local coordinates $\bm{\xi}$ of the isoparametric element, which are then a function of the joint's reference point $\bm{x}^i$. The shape functions are evaluated at these local coordinates:

\begin{equation}
w_j = N_j(\bm{\xi}(\bm{x}^i)),\qquad j \in\{1,\dots,n_n\}
\end{equation}

Equation~(\ref{eq:MPC}) will yield two (or three in 3D) linear equations per spring of each joint's pattern for all parts affected by the connection. The coefficients of all these coupling equations are put as row vectors into a couping matrix $\bm{G}$:

\begin{equation}
\begin{bmatrix}
\bm{g}^1(\bm{x}^1)\\ \vdots \\
\bm{g}^{n_j}(\bm{x}^{n_j}) 
\end{bmatrix} = \bm{G}(\bm{x}) \begin{bmatrix}
\bm{u}_m\\\bm{u}_c
\end{bmatrix} = \bm{0}
\end{equation}

$\bm{G}$ links all dependent slave connector displacements $\bm{u}_c$ to the master material displacements $\bm{u}_m$ of the parts.

The derivative of $\bm{G}$ uses the derivatives of the isoparametric shape functions:

\begin{equation}
\frac{dw_j}{dx_i} = \sum_k{\frac{dN_j}{d\xi_k}\frac{d\xi_k}{dx_i}}
\end{equation}

Lagrange multipliers $\bm{\lambda}_c$ are used to incorporate the coupling equations directly into the system matrix (see section~\ref{ss:system}). The Lagrange multipliers are then the constraint forces needed to keep the slave nodes at the desired positions.

\section{Sensitivities of objectives and constraints}
\label{sec:optimization}

The deterministic compliance (see section~\ref{ss:cdet}) and the compliance under failure of joints (see section~\ref{ss:cfail}) are studied as objective functions in this paper.

As constraint functions of the optimization, either one global, or several part-wise defined volume constraints (see section~\ref{ss:volc}), and in some cases additionally a minimum distance constraint (see section~\ref{ss:mindist}) are used.

\subsection{Objective function for deterministic compliance minimization}\label{ss:cdet}

The compliance is defined as:

\begin{equation}
c = \tilde{\bm{f}}^T \tilde{\bm{u}}
\end{equation}

Exploiting the structure of the augmented stiffness matrix from (\ref{eq:Kmatrix}) ,and expanding equation~(\ref{eq:KmKcG}), the compliance can be expressed as:

\begin{equation}
c = \bm{u}_m^T \bm{K}_m \bm{u}_m + \bm{u}_c^T \bm{K}_c \bm{u}_c + \begin{bmatrix}
\bm{u}_m^T&\bm{u}_c^T
\end{bmatrix} \bm{G}^T \bm{\lambda}_c + \bm{\lambda}_c^T \bm{G} \begin{bmatrix}
\bm{u}_m\\\bm{u}_c
\end{bmatrix}
\end{equation}

The linear coupling equations linking the displacements of the material nodes with the joint nodes sum up to zero:

\begin{equation}
\bm{G} \begin{bmatrix}
\bm{u}_m\\\bm{u}_c
\end{bmatrix} = \bm{0}
\end{equation}

Therefore, the last two summands do not contribute to the compliance (which means that the coupling equations do not store elastic energy). Hence, the compliance can be subdivided into the compliance contributions of every part plus the contributions of every joint:

\begin{subequations}\label{eq:ctotcmcc}
	\begin{align}
	c &= \bm{u}_m^T \bm{K}_m \bm{u}_m + \bm{u}_c^T \bm{K}_c \bm{u}_c\\
	&= \sum_{i=1}^{n_p}{\bm{u}_m^i}^T \bm{K}_m^i \bm{u}_m^i + \sum_{i=1}^{n_j}{\bm{u}_c^i}^T \bm{K}_c^i \bm{u}_c^i\\
	&= \sum_{i=1}^{n_p} c_m^i + \sum_{i=1}^{n_j} c_c^i\\
	&= c_m + c_c
	\end{align}
\end{subequations}

The total derivative of the compliance with respect to the material design variables is:

\begin{equation}
\frac{d c}{d \varrho_i} = \sum_j{\frac{\partial c}{\partial \hat{\varrho}_j}\frac{\partial \hat{\varrho}_j}{\partial \bar{\varrho}_j}\frac{\partial \bar{\varrho}_j}{\partial \tilde{\varrho}_j}\frac{\partial \tilde{\varrho}_j}{\partial \varrho_i}}\label{eq:dcdrho}
\end{equation}

The first term is obtained by adjoint sensitivity analysis~\cite{bendsoe_topology_2004}:

\begin{equation}
\frac{\partial c}{\partial \hat{\varrho}_j} = -\tilde{\bm{u}}^T\frac{\partial\tilde{\bm{K}}}{\partial \hat{\varrho}_j}\tilde{\bm{u}} = -\tilde{\bm{u}}^T\frac{\partial \bm{K}_m}{\partial \hat{\varrho}_j}\tilde{\bm{u}} = -\bm{u}^T_m\frac{\partial \bm{K}_m}{\partial \hat{\varrho}_j}\bm{u}_m\label{eq:dcdrhomod}
\end{equation}

Equation (\ref{eq:dcdrhomod}) utilizes that only the material part of $\tilde{\bm{K}}$ is dependent on the modified densities.

Using adjoint analysis for the derivative with respect to the joints' reference point coordinates yields:

\begin{equation}
\frac{d c}{d x_i} = -\tilde{\bm{u}}^T\frac{d\tilde{\bm{K}}}{d x_i}\tilde{\bm{u}}\label{eq:dcdx1}
\end{equation}

The augmented stiffness matrix is affected directly by the joints' positions through the coupling equations and indirectly through a change of the masks used for the modified densities:

\begin{subequations}
	\begin{align}
	\frac{d\tilde{\bm{K}}}{d x_i} &= \frac{\partial\tilde{\bm{K}}}{\partial x_i} + \sum_j{\frac{\partial \tilde{\bm{K}}}{\partial \hat{\varrho}_j}\frac{\partial \hat{\varrho}_j}{\partial x_i}}\\
	&= \frac{\partial \bm{G}}{\partial x_i} + \sum_j{\frac{\partial \bm{K}_m}{\partial \hat{\varrho}_j}\frac{\partial \hat{\varrho}_j}{\partial x_i}}\label{eq:dKmoddxi}
	\end{align}
\end{subequations}

Inserting (\ref{eq:dKmoddxi}) into equation (\ref{eq:dcdx1}) and using (\ref{eq:dcdrhomod}) yields:

\begin{equation}
\frac{d c}{d x_i} = -\tilde{\bm{u}}^T\frac{\partial \bm{G}}{\partial x_i}\tilde{\bm{u}} + \sum_j{\frac{\partial c}{\partial \hat{\varrho}_j}\frac{\partial \hat{\varrho}_j}{\partial x_i}}\label{eq:dcdx}
\end{equation}

The derivatives $\frac{\partial \hat{\varrho}_j}{\partial (\cdot)}$ used in equations (\ref{eq:dcdrho}) and (\ref{eq:dcdx}) are dependent on the type of the joints' material zones and are listed in Appendix~\ref{ap:rhomodderiv}.

\subsection{Objective function considering failure of joints}\label{ss:cfail}

To optimize a structure for minimum compliance in the case, that one or more joints fail, failure cases are defined. In each failure case, the stiffness of a unique combination of joints is degraded by multiplying it with a very low value, while all other joints remain intact with their full stiffness. Then, the maximum compliance for all failure cases is minimized.

The failure mode $m$ is defined by $m$ joints being considered as failed at the same time. Then, the total number $n_d$ of combinations is:

\begin{equation}
n_d = \begin{pmatrix}
n_j\\m
\end{pmatrix} = \frac{n_j!}{m!(n_j-m)!}
\end{equation}

The Kreisselmeier-Steinhauser (KS) formula~\cite{kreisselmeier_application_1983} is a conservative approximation of the max-operator and is used to combine the individual compliance values $c_d$ of the damage cases for failure mode $m$ into a single objective function:

\begin{equation}
c_{KS}^m = \frac{1}{\gamma}\log\left(\sum_d\exp(\gamma c_d^m)\right),\qquad d=1,\dots,n_d(m)
\end{equation}

The parameter $\gamma$ is a scaling factor of the KS function. The evaluation of the compliance for every failure scenario $c_d$ requires an own FE calculation. However, for compliance optimization, the worst case compliance of mode $m$ will always exceed all worst case compliances of the lower failure modes:

\begin{equation}
\max_d{c_d^m} > \max_d{c_d^{m-1}}, \qquad \forall m=1,\dots,n_j
\end{equation}

Therefore, only the failure scenarios from the highest studied failure mode need to be evaluated, when optimizing for the worst case. Since in practice the number of joints is relatively small, also the computational effort of considering failure of joints is small, compared to optimizations considering failure of the continuum structure~\cite{ambrozkiewicz_density-based_2020,jansen_topology_2013}.

The examples shown in section~\ref{sec:ex2D} and \ref{sec:ex3D} have a maximum number of $n_j = 4$ joints, which leads to $n_d = 4$ failure cases for failure mode $m=1$ and $n_d = 6$ failure cases for failure mode $m=2$.

\subsection{Volume constraints}\label{ss:volc}

A global volume constraint limits the total volume of the assembly to a maximum global volume fraction $k_g$. With $v_i$ being the volume of element $i$, the constraint function states:

\begin{subequations}
	\begin{align}
	h &= \frac{V}{V_0} -k_g \leq 0\\
	&= \frac{\sum_i \hat{\varrho}_i v_i}{\sum_i v_i} -k_g \leq 0\label{eq:GVC}
	\end{align}
\end{subequations}

The volume constraint can also be applied on a per-part basis to have control over how the material is distributed among the parts. The evaluation of the volumes is then performed over a set $\mathbb{L}_j$ containing the element indices of part $j$. The maximum volume fraction per part can then be limited by a value $k_l$:

\begin{align}
h_j &= \frac{\sum_{i\in\mathbb{L}_j} \hat{\varrho}_i v_i}{\sum_{i\in\mathbb{L}_j} v_i} -k_l \leq 0,\qquad j = 1,\dots,n_p\label{eq:LVC}
\end{align}

The derivatives with respect to the material and position design variables are given in Appendix~\ref{ap:volcderiv}.

\subsection{Minimum distance constraint}\label{ss:mindist}

A minimum distance between individual joints may be needed to avoid overlapping and to ensure a minimum clearance of the fastener elements in the assembly.

The squared distance $s^{ij}$ between the reference points of joints $i$ and $j$ is:

\begin{equation}
s^{ij} = {d^{ij}}^2 = (\bm{x}^j - \bm{x}^i)^T(\bm{x}^j - \bm{x}^i)
\end{equation}

The pairwise distances $d^{ij} = \sqrt{s^{ij}}$ should not fall below a desired value $d_0$. Either each distance may be constrained by an own constraint function or a single aggregated function constraining the minimal distance can be used:

\begin{equation}
h_{d}(\bm{x}) = d_0 - \sqrt{\min_{i \neq j}s^{ij}} \leq 0\label{eq:hmindist}
\end{equation}

For gradient based optimization, a differentiable approximation for the min-operator in equation (\ref{eq:hmindist}) is needed. A common choice for approximating the max-operator is the p-norm, which approaches the real maximum absolute value for increasing $p$ values. Here, a reciprocal mapping is applied to the squared distances $s^{ij}$ to alter the min-problem to a max-problem while maintaining strictly positive values. Then the p-norm can be used and the end result needs to be inverted again:

\begin{equation}
h_{d,agg}(\bm{x}) = d_0 - \left(\sum_i\sum_{j < i}\left(s^{ij}+\epsilon\right)^{-p}\right)^{-\frac{1}{2p}}\leq 0\label{eq:hmdpnorm}
\end{equation}

Equation (\ref{eq:hmdpnorm}) takes into account, that $s^{ij} = s^{ji}$, such that the number of summands is halved and therefore the quality of the approximation is improved. Also, a parameter $\epsilon$ with a small positive value is introduced to counter the singularity that would otherwise exist, if points coincide. In this paper, a value of 0.01 times the element edge length is used for $\epsilon$ and $p$ is set to a value of 8. The gradient of the aggregated function is obtained by direct differentiation and is given in Appendix~\ref{ap:mindistderiv}.

\section{Examples in 2D}\label{sec:ex2D}

\begin{figure}
	\centering
	\subfloat[Single part design\label{fig:DS2D_full}]{\includegraphics[trim={50, 110, 35, 100},clip,width=0.475\linewidth]{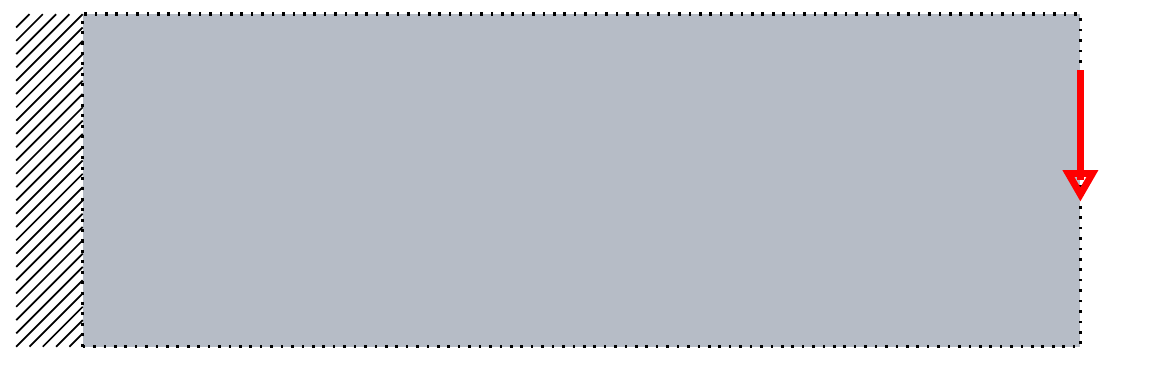}}\hfill
	\subfloat[Two parts design with area of overlap\label{fig:DS2D_layers}]{\includegraphics[trim={50, 100, 35, 90},clip,width=0.475\linewidth]{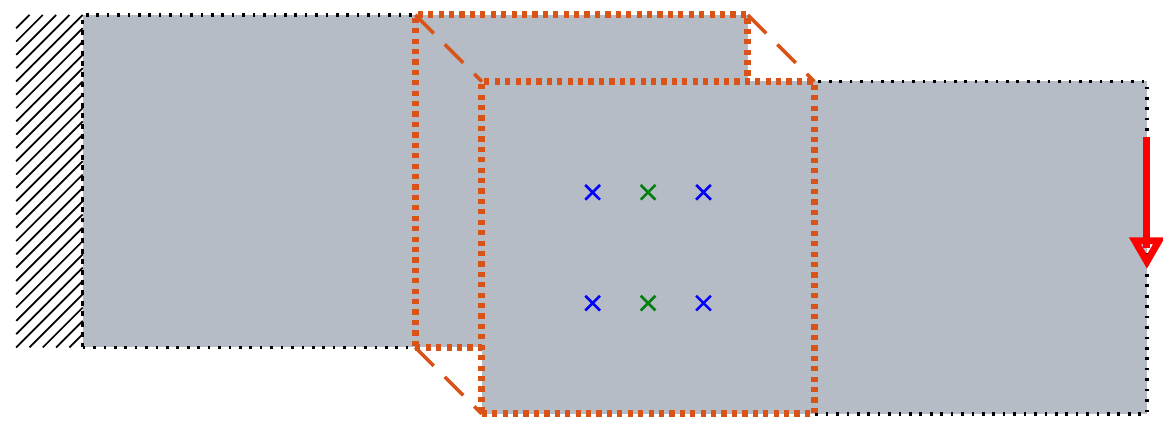}}
	\caption{Design domains for the 2D examples with boundary conditions, loads and initial locations of the joints (green, blue).}
	\label{fig:DS2D}
\end{figure}

For the 2D examples, a cantilever beam is optimized for minimum compliance under a volume constraint. A single part design will serve as a reference, while the other designs will consist of two separate parts connected by joints. The design domains, boundary conditions, and loads are shown in figure~\ref{fig:DS2D}.

The design space of the reference design is discretized by $300 \times 100$ finite elements. The two parts design has the same outer dimensions, while each part has a design domain of $200 \times 100$ elements. Therefore, an overlapping area of $100 \times 100$ elements is present, which bounds the space where the joints can be located.

The initial positions of the joints' reference points for the examples involving two joints are shown in green in figure~\ref{fig:DS2D_layers}. The positions marked in blue refer to the initial positions for the examples involving four joints. Parameters used for the optimization are listed in Appendix~\ref{ap:2Dparams}.

\subsection{Reference design}

\begin{figure}
	\centering
	\includegraphics[trim={50, 105, 35, 95},clip,width=0.5\linewidth]{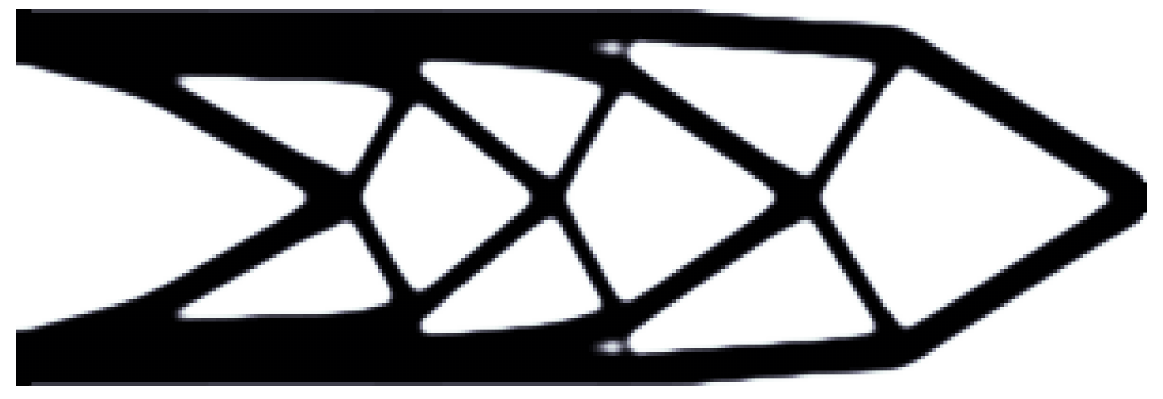}
	\caption{Reference design for the 2D examples.}
	\label{fig:DetRef}
\end{figure}

The reference design employs a global volume constraint limiting the global volume fraction to $k_g = 0.4$. The final design has a compliance of $c = 210.19$ and is shown in figure~\ref{fig:DetRef}.

\subsection{Design with 2 joints}\label{ss:2J}

Two different joint types are investigated here: Spot welds, which only require a circular material zone, and bolts, which additionally require a mounting hole through both mating parts.

The load transfer is modeled by concentric patterns of springs. The resultant stiffness of the spring patterns is the same for both connection types and has the value $k_c = 10$. The spring patterns, together with the material and force transfer zones, are shown in figure~\ref{fig:Patterns2D}. For the spot welds, a material radius of 8 is used, together with a force transfer zone radius of 4. For the bolted design, the material zone has an outer radius of 10, the hole has a radius of 4. The radii for the springs are 6 and 8.

For the assembled designs, a part-wise volume constraint of $k_l = 0.3$  is employed for both parts, yielding the same total mass as the reference design.

\begin{figure}
	\centering
	\subfloat[Left part\label{fig:Det2Jl}]{\includegraphics[trim={50, 105, 145, 95},clip,width=0.25\linewidth]{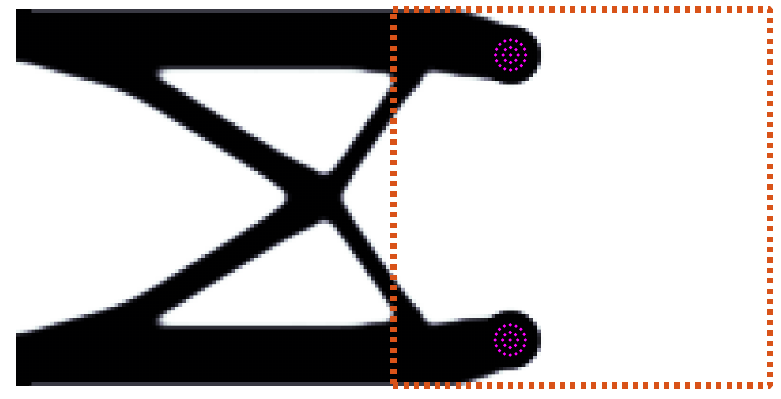}}
	\subfloat[Right part\label{fig:Det2Jr}]{\includegraphics[trim={160, 105, 35, 95},clip,width=0.25\linewidth]{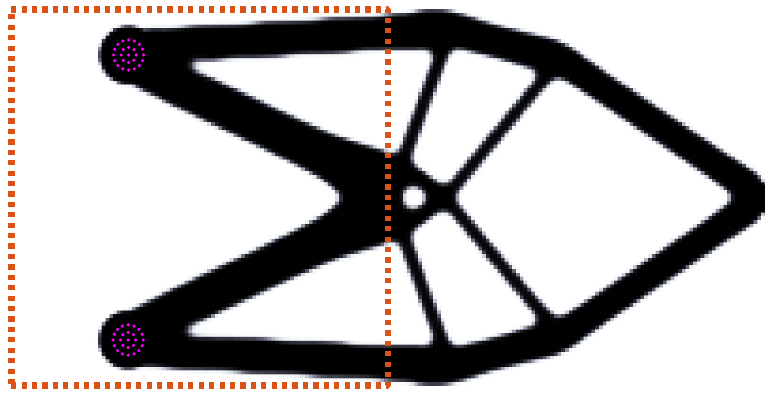}}\\
	\subfloat[Assembly\label{fig:Det2Jb}]{\includegraphics[trim={50, 105, 35, 95},clip,width=0.5\linewidth]{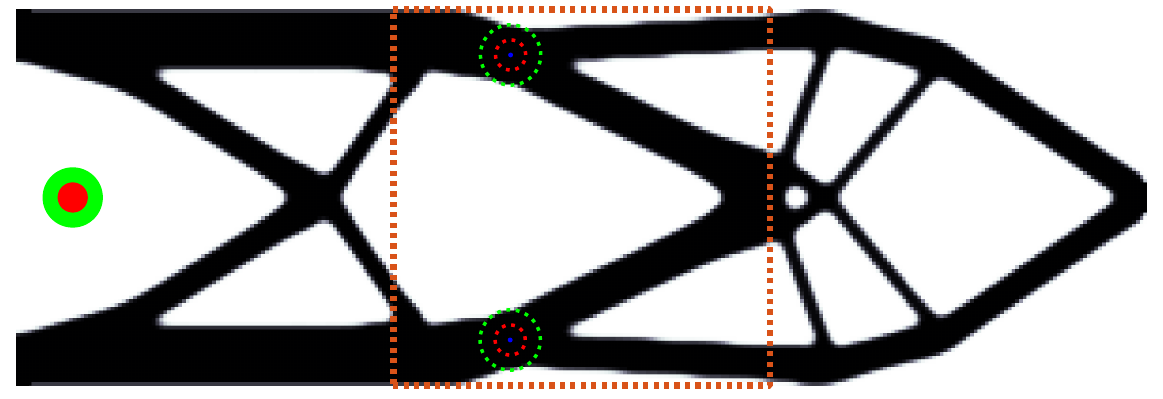}}
	\caption{Results for 2 spot welds.}
	\label{fig:Det2J}
\end{figure}

\begin{figure}
	\centering
	\subfloat[Left part\label{fig:Det2JRingl}]{\includegraphics[trim={50, 105, 145, 95},clip,width=0.25\linewidth]{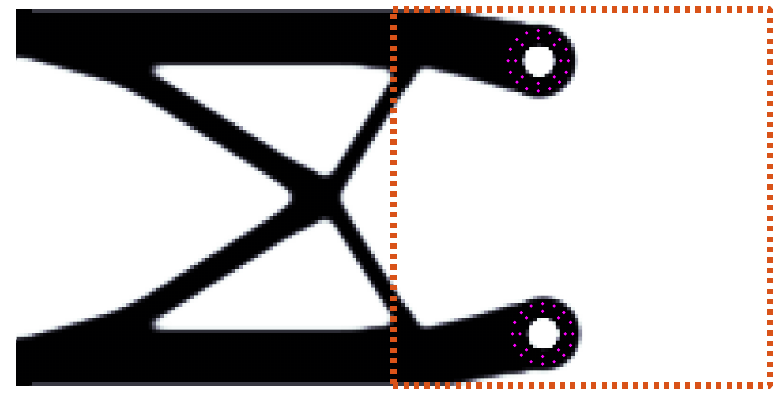}}
	\subfloat[Right part\label{fig:Det2JRingr}]{\includegraphics[trim={160, 105, 35, 95},clip,width=0.25\linewidth]{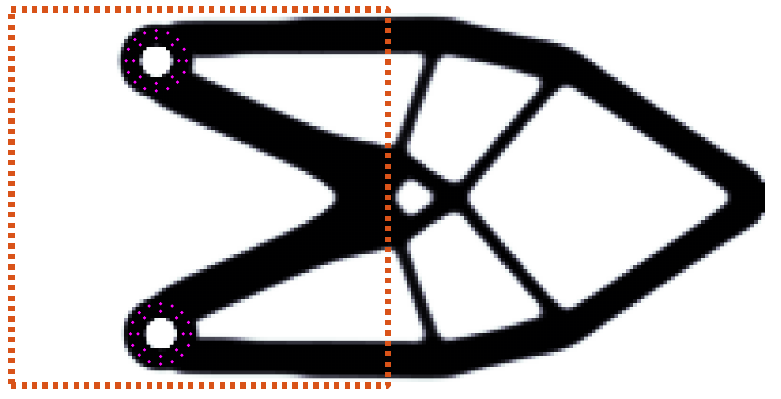}}\\
	\subfloat[Assembly\label{fig:Det2JRingb}]{\includegraphics[trim={50, 105, 35, 95},clip,width=0.5\linewidth]{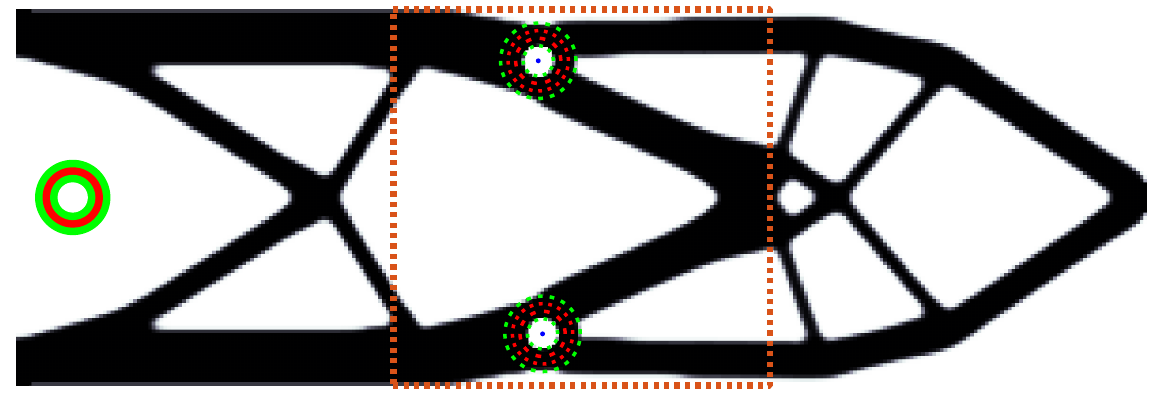}}
	\caption{Results for 2 bolts.}
	\label{fig:Det2JRing}
\end{figure}

The final designs for the spot welded and bolted variants are shown in figure~\ref{fig:Det2J} and figure~\ref{fig:Det2JRing}, respectively. The results share the same topology and differ only slightly. In both cases, the joints are placed on the main load paths on the thick upper and lower beams of the structure.

Numerical values for the compliance $c$ are given in table~\ref{tab:tab2J} and reveal, that the assembled designs only suffer from a small penalty of about 6\% and 7\% compared to the one-piece design. The spot weld design is slightly stiffer than the bolted design since the latter has a weakening hole in the load path and also the size of the enforced material zone is larger, making it harder to integrate it into the structure.

In table~\ref{tab:tab2J} the total compliance $c$ is further split into the compliance contribution of the material $c_m$ and of the joints $c_c$, according to equation (\ref{eq:ctotcmcc}). With the parameters used here, the joint compliance is just a very small fraction of the total compliance, such that the optimization problem is dominated by the material distribution and the joint locations, and not by the modeling of the force transfer in the joints.

\begin{table}
	\centering
	\caption{Compliance values for the 2 joint design.}
	\label{tab:tab2J}
	\begin{tabular}{lrrr}
		\hline\noalign{\smallskip}
		Configuration & total $c$ & material $c_m$ & joint $c_c$ \\
		\noalign{\smallskip}\hline\noalign{\smallskip}
		Reference (fig.~\ref{fig:DetRef})& 210.19 & 210.19 & 0\\
		2 spots (fig.~\ref{fig:Det2J})& 223.00& 221.80 &1.20\\
		2 bolts (fig.~\ref{fig:Det2JRing})& 225.41 & 224.24 & 1.18\\
		\noalign{\smallskip}\hline
	\end{tabular}
\end{table}

\subsection{Design with 4 joints}\label{ss:4J}

When increasing the number of joints, the minimum distance constraint from equation~(\ref{eq:hmdpnorm}) needs to be included in the optimization task. Otherwise, joints may coincide, which leads to an unphysical joint layout and thus should be avoided. Without modifying the optimization task, the results for a four joint design would look the same as the results from section~\ref{ss:2J} with just two joints.

\begin{figure}
	\centering
	\subfloat[Left part\label{fig:Det4JRingl}]{\includegraphics[trim={50, 105, 145, 95},clip,width=0.25\linewidth]{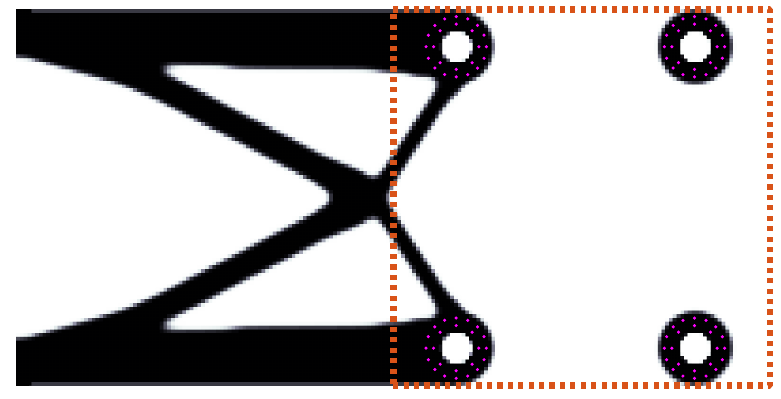}}
	\subfloat[Right part\label{fig:Det4JRingr}]{\includegraphics[trim={160, 105, 35, 95},clip,width=0.25\linewidth]{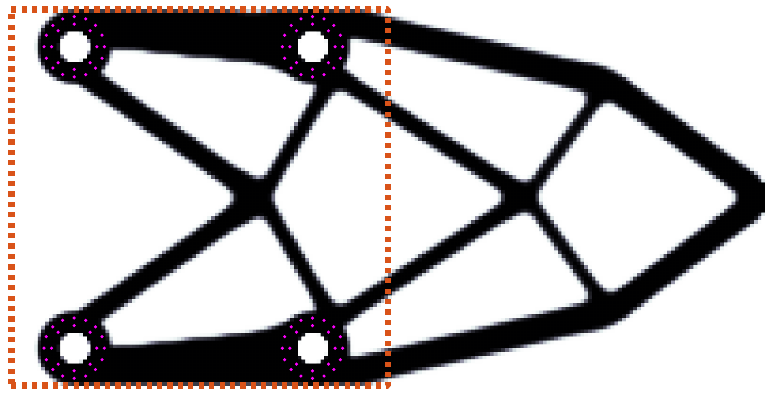}}\\
	\subfloat[Assembly\label{fig:Det4JRingb}]{\includegraphics[trim={50, 105, 35, 95},clip,width=0.5\linewidth]{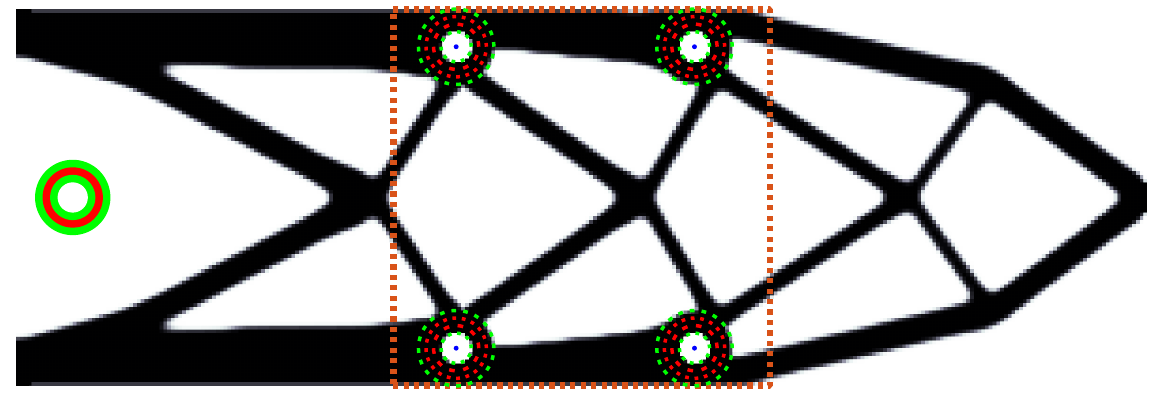}}
	\caption{Results for 4 bolts with minimum distance constraint.}
	\label{fig:Det4JRing}
\end{figure}

Since the two joint design is already close to the single part reference design in terms of performance, additional joints are virtually unnecessary. When employing a minimum distance constraint with $d_0 = 20$ for the bolted design, the result from figure~\ref{fig:Det4JRing} is obtained. The two rightmost joints do not transfer any load between the parts. These unused joints come with unused material, therefore the structural performance is worse, than for the two bolt design, as shown in table~\ref{tab:tab4J}.

\begin{table}
	\centering
	\caption{Compliance values for the bolted design.}
	\label{tab:tab4J}
	\begin{tabular}{lrrr}
		\hline\noalign{\smallskip}
		Configuration & total $c$ & material $c_m$ & joint $c_c$ \\
		\noalign{\smallskip}\hline\noalign{\smallskip}
		Reference (fig.~\ref{fig:DetRef})& 210.19 & 210.19 & 0\\
		2 bolts (fig.~\ref{fig:Det2JRing})& 225.41 & 224.24 & 1.18\\
		4 bolts (fig.~\ref{fig:Det4JRing})& 226.63 & 224.90 & 1.73\\
		\noalign{\smallskip}\hline
	\end{tabular}
\end{table}

The unused joints are placed at a location where the waste of material is minimal. The structure in figure~\ref{fig:Det4JRingb} shows a strong similarity with the reference design in figure~\ref{fig:DetRef} which also has small holes at the same locations, where the holes for the unused joints are placed.

\subsection{Design with 4 joints considering failure}

\begin{figure}
	\centering
	\subfloat[Left part\label{fig:FS4JRingl}]{\includegraphics[trim={50, 105, 145, 95},clip,width=0.25\linewidth]{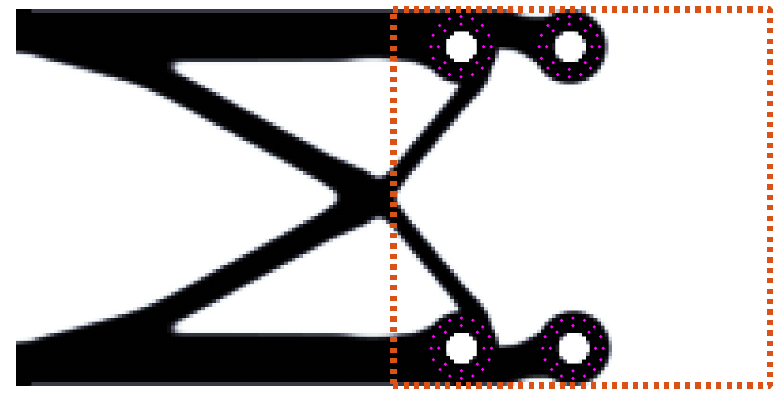}}
	\subfloat[Right part\label{fig:FS4JRingr}]{\includegraphics[trim={160, 105, 35, 95},clip,width=0.25\linewidth]{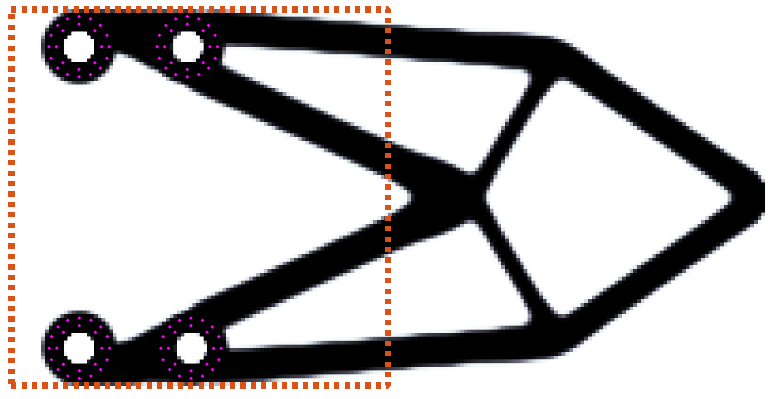}}\\
	\subfloat[Assembly\label{fig:FS4JRingb}]{\includegraphics[trim={50, 105, 35, 95},clip,width=0.5\linewidth]{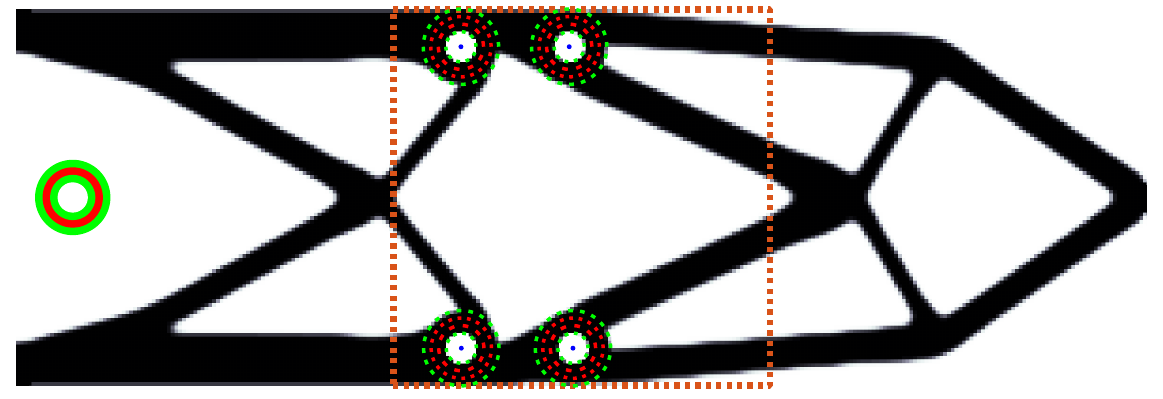}}
	\caption{Results for 4 bolts with minimum distance constraint, fail-safe design.}
	\label{fig:FS4JRing}
\end{figure}

When considering the failure of single joints, it is even more important to employ a minimum distance constraint, otherwise, joints will overlap in pairs of two for ideal (but unphysical) redundancy. When enforcing a minimum distance, the joints are pushed further apart. For $d_0 = 20$, the result in figure~\ref{fig:FS4JRing} is obtained, which now has no unused joints.

\begin{table}
	\centering
	\caption{Table with values for the 4 bolt design.}
	\label{tab:tab4JFS}
	\begin{tabular}{lrr}
		\hline\noalign{\smallskip}
		Configuration & nominal $c$ & max $c_d^1$\\
		\noalign{\smallskip}\hline\noalign{\smallskip}
		4 bolts (fig.~\ref{fig:Det4JRing})& 226.63 & 7903.42\\
		4 bolts, fail-safe (fig.~\ref{fig:FS4JRing}) & 231.43 & 239.29\\
		\noalign{\smallskip}\hline
	\end{tabular}
\end{table}

Table~\ref{tab:tab4JFS} lists the compliance values for the nominal, undamaged case as well as for the worst-case failure of a single joint (failure mode $m=1$).

Without considering damage, the fail-safe design is more compliant than the design from section~\ref{ss:4J} with four joints being explicitly optimized for the undamaged case, even though latter has unused joints. However, when considering single failure, the fail-safe design suffers only from a small loss of stiffness, while the deterministic design is in the worst case left with only one load-bearing joint, causing a severe drop of stiffness.

\section{Examples in 3D}\label{sec:ex3D}

\begin{figure}
	\centering
	\subfloat[Contact surface (orange), boundary condition edges (purple) and initial joint positions (blue)\label{fig:DS3D_DS}]{\includegraphics[trim={80, 60, 65, 75},clip,width=0.475\linewidth]{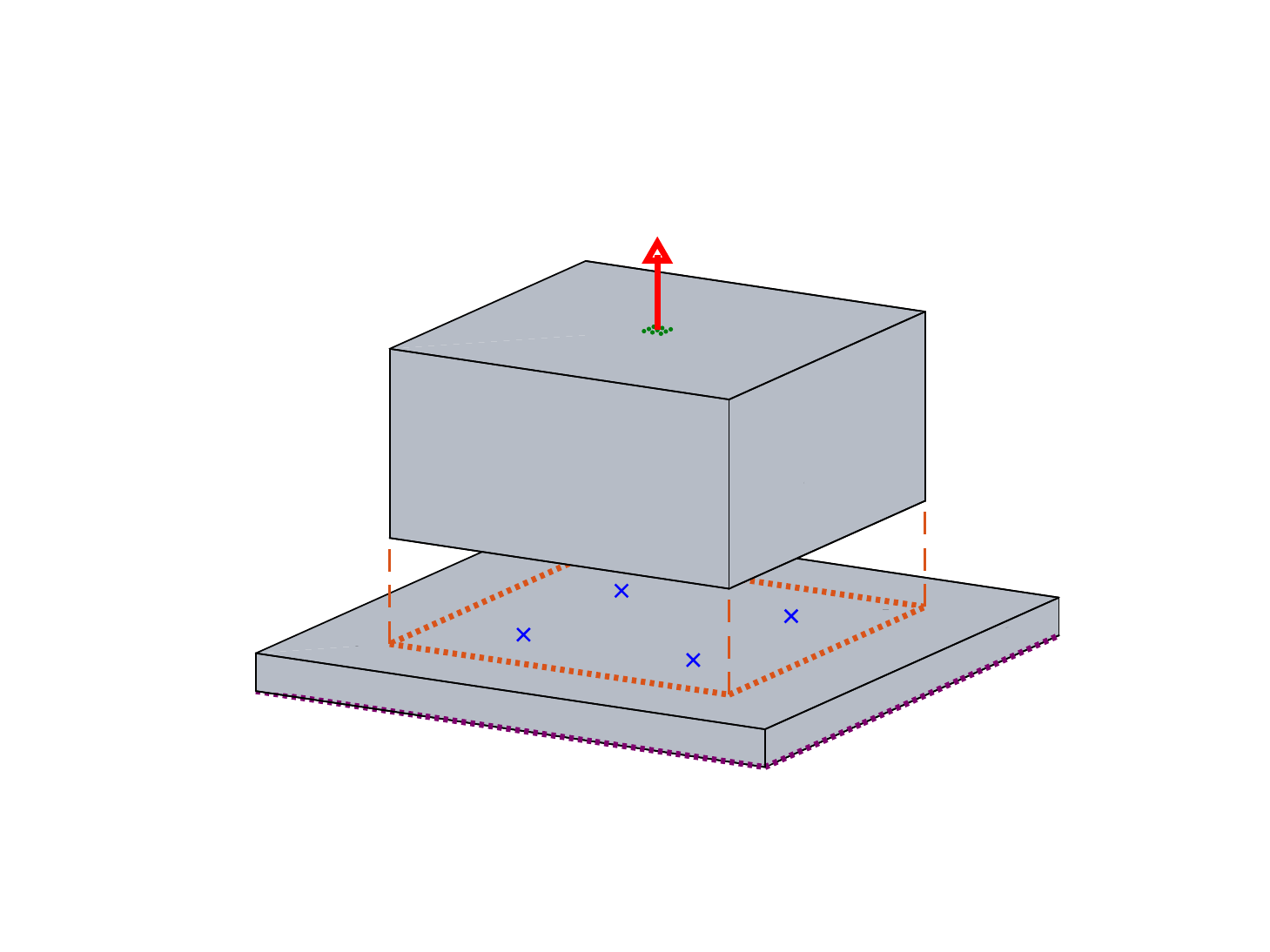}}\hfill
	\subfloat[Load distribution at the top\label{fig:DS3D_loads}]{\includegraphics[trim={50, 45, 35, 90},clip,width=0.475\linewidth]{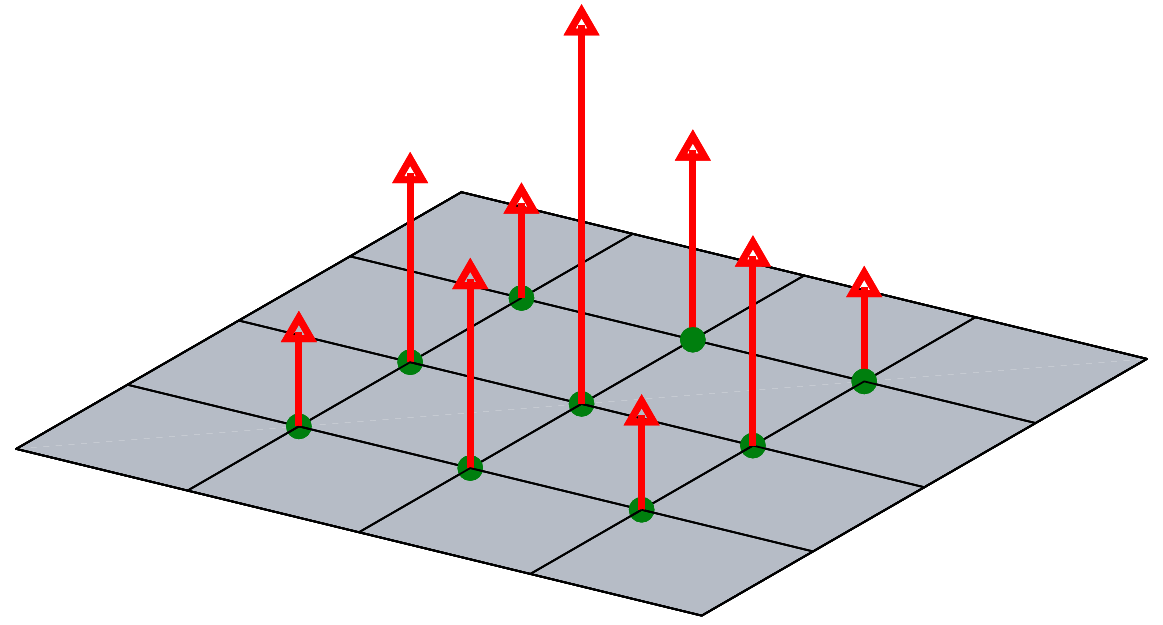}}
	\caption{Design space and loading for the 3D examples.}
	\label{fig:DS3D}
\end{figure}

For the 3D example, a plate and a bracket connected by four joints and subjected to a pulling load are studied. The design spaces are shown in figure~\ref{fig:DS3D_DS}. The plate is fixed along the lower edges in the upward direction and at one corner also in the in-plane directions. The load points upwards and is distributed on the nine central nodes of the upper surface of the bracket, as shown in figure~\ref{fig:DS3D_loads}.

The bracket is discretized by $40 \times 40 \times 20$ finite elements, the plate by $60 \times 60 \times 4$ elements. Optimization parameters are listed in Appendix~\ref{ap:3Dparams}.

\begin{figure}
	\centering
	\includegraphics[trim={150, 40, 135, 25},clip,width=0.2\linewidth]{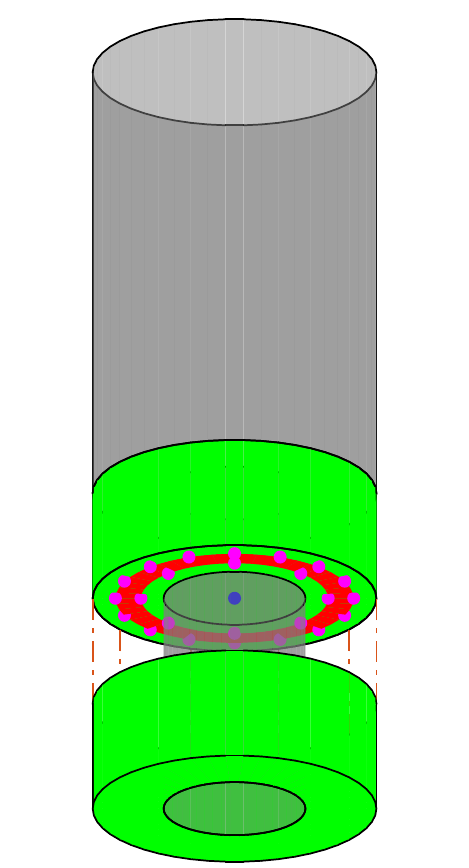}
	\caption{NDS Solid zones (green, red), clearance (gray), reference point (blue) and spring pattern (magenta) used in the 3D examples.}
	\label{fig:Pattern3D}
\end{figure}

A material zone with the shape of a hollow cylinder is used for every joint, as shown in figure~\ref{fig:Pattern3D}. The outer material radius is 5, the hole has a radius of 2.5. The height of the material zone is 4 on the bracket and also 4 on the plate side. Above the material zone, a clearance zone, granting accessibility for bolts mounted from the top, is considered. The top clearance area only affects the bracket and has an outer radius of 5. The spring pattern for the force transfer zone shown in figure~\ref{fig:Pattern3D} is equivalent to the pattern in figure~\ref{fig:RingPattern2D} with radii 3.3 and 4.2, and is located on the contact surface of both parts.

While the bracket undergoes a three-dimensional topology optimization, the plate is optimized for an extruded ``2.5D'' design, with no density variation along its thickness. A part-wise volume constraint is employed with a volume fraction of 15\% for the bracket and 40\% for the plate.

\begin{figure}
	\centering
	\subfloat[Deterministic result\label{fig:FS0}]{\includegraphics[width=0.3\linewidth]{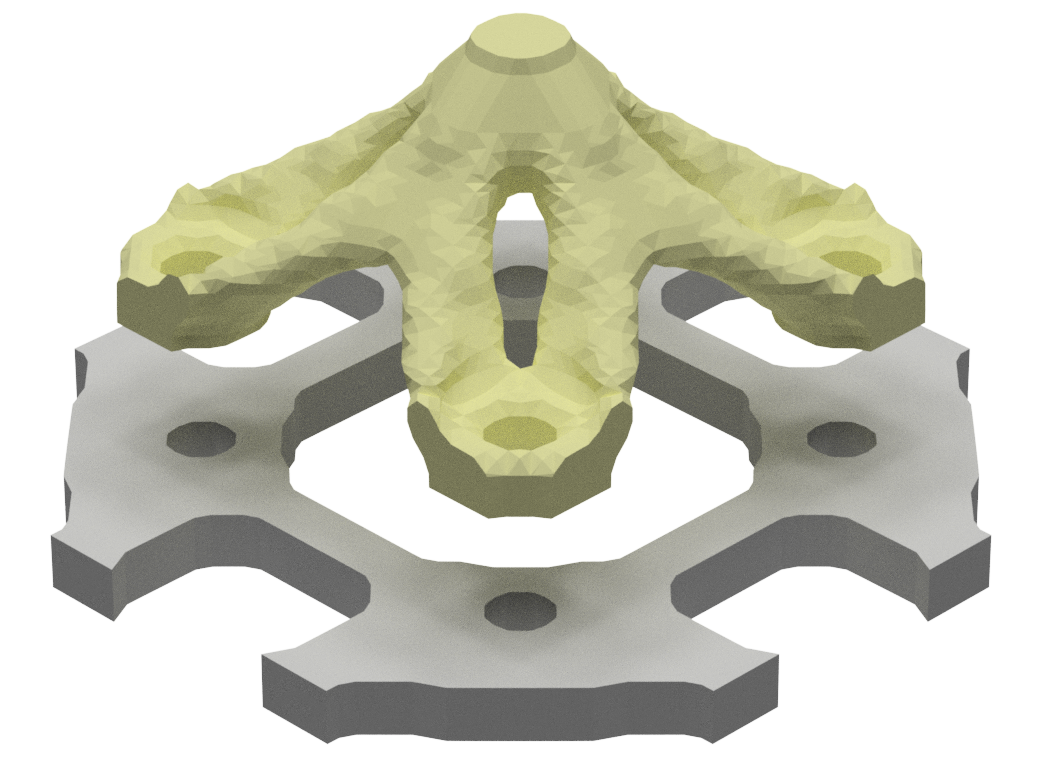}}\\
	\subfloat[Result considering single failure\label{fig:FS1}]{\includegraphics[width=0.3\linewidth]{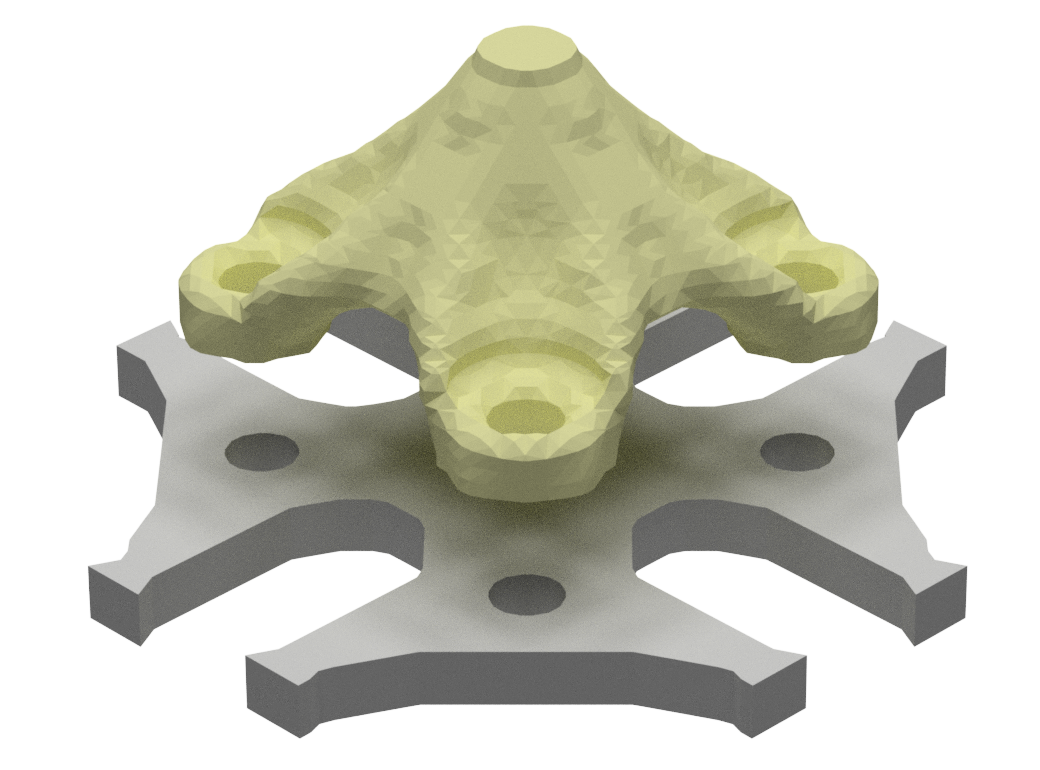}}\\
	\subfloat[Result considering double failure\label{fig:FS2}]{\includegraphics[width=0.3\linewidth]{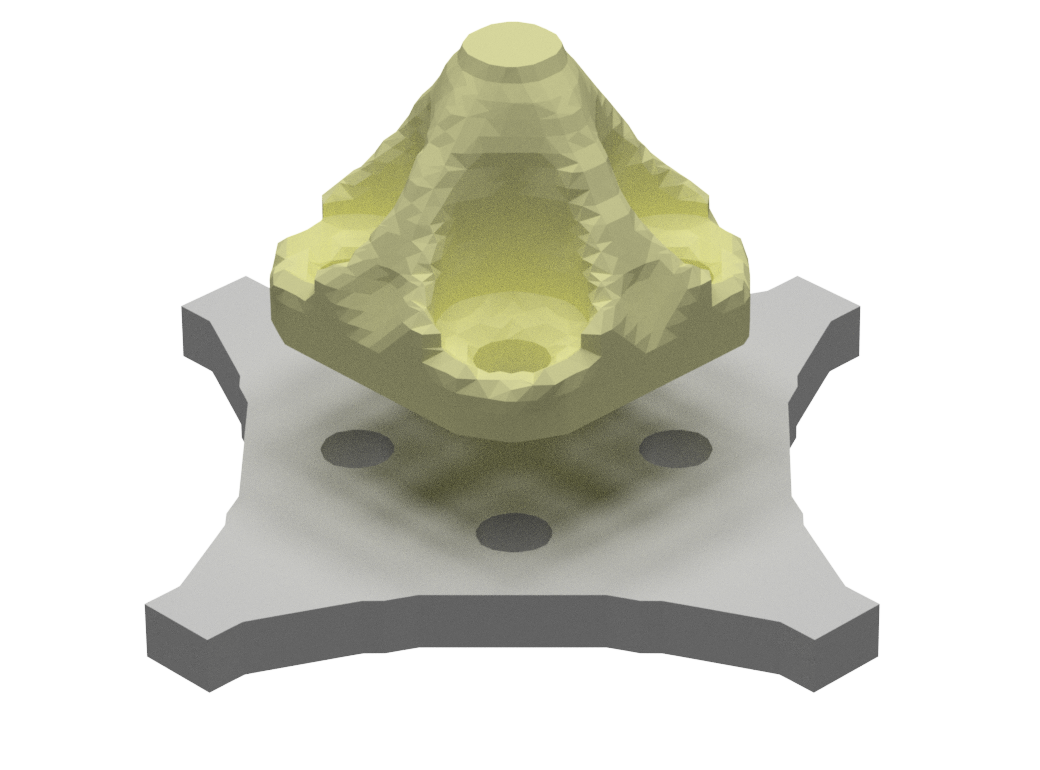}}
	\caption{Results for the 3D example.}
	\label{fig:Result3D}
\end{figure}

Results for the 3D examples are depicted in figure~\ref{fig:Result3D}. The shown parts are obtained by extracting an isosurface of the density field from the finite element mesh at $\hat{\bm{\varrho}} = 0.5$, yielding a tessellated surface of the parts. The parts are shown with an offset, such that also the plate can be seen.

The result for the deterministic optimization without consideration of failure is shown in figure~\ref{fig:FS0}. When considering that both parts carry the load as an assembly, the mounting points are moved as far out as possible, such that the bracket has the maximum width. In contrast to that, if the bracket would be optimized as a single part under a pulling load, a compact rod would be obtained. If mounted on a plate, the plate would then be loaded unfavorably at its most compliant location in the middle. This example demonstrates the importance of optimizing assemblies as a whole, to counter the effect that a single optimized part may introduce suboptimal interface loads to surrounding parts.

Figures~\ref{fig:FS1} and \ref{fig:FS2} show the resulting designs, when considering failure of a single joint ($m=1$) or two joints at once ($m=2$), respectively. For increased robustness with respect to joint failures, the individual joints are grouped closer together. However, the nominal compliance for the undamaged case suffers an increasing penalty, when comparing the numerical values for the compliance given in table~\ref{tab:tab3D}. The rows in table~\ref{tab:tab3D} refer to the failure mode considered in the optimization. The columns show the compliance values for the nominal, undamaged case as well as the worst-case damages for a single and a double joint failure.

\begin{table}
	\centering
	\caption{Compliance values for the 4 bolt 3D design.}
	\label{tab:tab3D}
	\begin{tabular}{lrrr}
		\hline\noalign{\smallskip}
		Configuration & nominal $c$ & max $c_d^1$ & max $c_d^2$\\
		\noalign{\smallskip}\hline\noalign{\smallskip}
		Deterministic & 2.0893 & 4.1229 & 25.0710\\
		Single failure & 2.7796 & 3.7467 & 13.4129\\
		Double failure & 4.3491 & 4.8565 & 6.9133\\
		\noalign{\smallskip}\hline
	\end{tabular}
\end{table}

\section{Conclusion and outlook}\label{sec:outlook}

An efficient method for the simultaneous optimization of parts and their connection inside assemblies has been presented. The 2D and 3D examples show the applicability and potentials of the method. By considering material and clearance zones for joint mountability already inside the numerical models, a consistency between the simulated and manufactured part geometry is obtained. Additionally considering the failure of joints allows for trading off nominal performance in favor of residual performance under failed conditions, which may significantly be improved at a reasonable computational cost.

This paper focused on ``point-type'' connections, like spot welds, rivets, or bolts, for which the location is described by a single reference point per joint. The method is also applicable to ``line-type'' connections like e.g. weld seams or glue lines. An open polygon or an open spline curve would be the reference curve for these types of connections. The position design variables are in this case the coordinates of the polygon's edges or the spline's control points. Again, a pattern of springs is attached to the reference curve to model the size of the force transfer zone. A dilated version of the reference curve can be used to incorporate the minimum material zone.


\section*{Acknowledgments}
Financial support of the German Ministry for Economic Affairs and Energy (project REGIS, funding reference 20W1708E) and the German Research Foundation (reference number KR 4914/3-1) is acknowledged.

\subsection*{Conflict of interest}

On behalf of all authors, the corresponding author states that there is no conflict of interest.

\appendix

\section{Sensitivities of the modified densities}\label{ap:rhomodderiv}

\subsection{Derivatives of a single mask}\label{ap:maskderiv}

The entries of a single mask vector are obtained via equation~(\ref{eq:maskval}). The derivatives with respect to the used reference point coordinates are:

\begin{subequations}
	\begin{align}
	\frac{\partial\psi^i_j}{\partial x^i} &= -\frac{\alpha x^i}{r_m^2}\sech^2{\left((\alpha E(\bm{c}^j - \bm{x}^i)\right)}\\
	\frac{\partial \psi^i_j}{\partial y^i} &= -\frac{\alpha y^i}{r_m^2}\sech^2{\left((\alpha E(\bm{c}^j - \bm{x}^i)\right)}
	\end{align}
\end{subequations}

\subsection{Derivatives for the introduction of holes}\label{ap:holederiv}

To introduce holes in the structure, equation~(\ref{eq:maskhole}) is used. The sensitivities with respect to the projected densities and the hole's center location at the reference point of joint $i$ are:

\begin{subequations}
	\begin{align}
	\frac{\partial\hat{\varrho}_i}{\partial\bar{\varrho}_i} &= \tensor[^-]{\psi}{_i}\\
	\frac{\partial\hat{\bm{\varrho}}}{\partial x^i} &= \bar{\bm{\varrho}} \circ \frac{\partial\tensor[^-]{\bm{\psi}}{}}{\partial x^i} = \bar{\bm{\varrho}} \circ \frac{\partial\tensor[^-]{\bm{\psi}}{^i}}{\partial x^i}\prod_{j\neq i}\tensor[^-]{\bm{\psi}}{^j}\\
	\frac{\partial\hat{\bm{\varrho}}}{\partial y^i} &= \bar{\bm{\varrho}} \circ \frac{\partial\tensor[^-]{\bm{\psi}}{}}{\partial y^i} = \bar{\bm{\varrho}} \circ \frac{\partial\tensor[^-]{\bm{\psi}}{^i}}{\partial y^i}\prod_{j\neq i}\tensor[^-]{\bm{\psi}}{^j}
	\end{align}
\end{subequations}

\subsection{Derivatives for the introduction of material}\label{ap:circderiv}

To add circular or cylindrical features of material to a structure, equation~(\ref{eq:maskcirc}) is used. The derivatives are in this case:

\begin{subequations}
	\begin{align}
	\frac{\partial\hat{\varrho}_i}{\partial\bar{\varrho}_i} &= \tensor[^+]{\psi}{_i}\\
	\frac{\partial\hat{\bm{\varrho}}}{\partial x^i} &= (\bar{\bm{\varrho}}-\bm{I}) \circ \frac{\partial\tensor[^+]{\bm{\psi}}{}}{\partial x^i} = (\bar{\bm{\varrho}}-\bm{I}) \circ \frac{\partial\tensor[^+]{\bm{\psi}}{^i}}{\partial x^i}\prod_{j\neq i}\tensor[^+]{\bm{\psi}}{^j}\\
	\frac{\partial\hat{\bm{\varrho}}}{\partial y^i} &= (\bar{\bm{\varrho}}-\bm{I}) \circ \frac{\partial\tensor[^+]{\bm{\psi}}{}}{\partial y^i} = (\bar{\bm{\varrho}}-\bm{I}) \circ \frac{\partial\tensor[^+]{\bm{\psi}}{^i}}{\partial y^i}\prod_{j\neq i}\tensor[^+]{\bm{\psi}}{^j}
	\end{align}
\end{subequations}

\subsection{Derivatives for the introduction of material rings}\label{ap:ringderiv}

The derivatives of equation~(\ref{eq:maskring}) are:

\begin{subequations}
	\begin{align}
	\frac{\partial\hat{\varrho}_i}{\partial\bar{\varrho}_i} &= \tensor[^+]{\psi}{_i} \tensor[^-]{\psi}{_i}\\
	\frac{\partial\hat{\bm{\varrho}}}{\partial x^i} &= \frac{\partial\tensor[^-]{\bm{\psi}}{}}{\partial x^i}+(\bar{\bm{\varrho}}-\bm{I}) \circ \left(\frac{\partial\tensor[^+]{\bm{\psi}}{}}{\partial x^i}\circ\tensor[^-]{\bm{\psi}}{}+ \tensor[^+]{\bm{\psi}}{}\circ\frac{\partial\tensor[^-]{\bm{\psi}}{}}{\partial x^i}\right)\\
	\frac{\partial\hat{\bm{\varrho}}}{\partial y^i} &= \frac{\partial\tensor[^-]{\bm{\psi}}{}}{\partial y^i}+(\bar{\bm{\varrho}}-\bm{I}) \circ \left(\frac{\partial\tensor[^+]{\bm{\psi}}{}}{\partial y^i}\circ\tensor[^-]{\bm{\psi}}{}+ \tensor[^+]{\bm{\psi}}{}\circ\frac{\partial\tensor[^-]{\bm{\psi}}{}}{\partial y^i}\right)
	\end{align}
\end{subequations}

\section{Sensitivities of the constraint functions}

\subsection{Derivatives of the volume constraints}\label{ap:volcderiv}

The derivatives of the part-wise volume constraint from equation~(\ref{eq:LVC}) with respect to the material and position design variables are:

\begin{align}
\frac{d h_k}{d \varrho_i} &= \sum_j{\frac{\partial h_k}{\partial \hat{\varrho}_j}\frac{\partial \hat{\varrho}_j}{\partial \bar{\varrho}_j}\frac{\partial \bar{\varrho}_j}{\partial \tilde{\varrho}_j}\frac{\partial \tilde{\varrho}_j}{\partial \varrho_i}}\\
\frac{d h_k}{d x_i} &= \sum_j{\frac{\partial h_k}{\partial \hat{\varrho}_j}\frac{\partial \hat{\varrho}_j}{\partial x_i}}
\end{align}

with:

\begin{equation}
\frac{\partial h_k}{\partial \hat{\varrho}_j} = \left\{{
	\begin{aligned}
	0,\qquad j \notin \mathbb{L}_k\\
	\frac{v_j}{\sum_{i\in\mathbb{L}_k} v_i},\qquad j \in \mathbb{L}_k
	\end{aligned}}
\right.
\end{equation}

\subsection{Derivatives of the minimum distance constraint}\label{ap:mindistderiv}

The gradient of the aggregated minimum distance constraint is obtained by direct differentiation of equation~(\ref{eq:hmdpnorm}):

\begin{subequations}
	\begin{gather}
	\frac{d h_{d,agg}}{d x_k} = {\scriptstyle -\frac{1}{2}\left(\sum\sum\left(s^{ij}+\epsilon\right)^{-p}\right)^{-\frac{1}{2p}-1}\left(\sum\sum\left(s^{ij}+\epsilon\right)^{-p-1} \frac{d s^{ij}}{d x_k}\right)}\\
	\begin{align}
	\frac{ds^{ij}}{d x^i} &= -2 (x^j-x^i), &\frac{ds^{ij}}{d y^i} &= -2 (y^j-y^i),\\
	\frac{ds^{ij}}{d x^j} &= 2 (x^j-x^i), &\frac{ds^{ij}}{d y^j} &= 2 (y^j-y^i)
	\end{align}
	\end{gather}
\end{subequations}

\section{Optimization and model parameters}

\subsection{Parameters for 2D examples}\label{ap:2Dparams}

\centering
\begin{tabular}{lr}
	\hline\noalign{\smallskip}
	Parameter & Value\\
	\noalign{\smallskip}\hline\noalign{\smallskip}
	Element size & $1 \times 1$\\
	SIMP penalty exponent $p$ & 3\\
	Filter radius $r$ & 4\\
	Young's modulus $E_0$ & 1\\
	Poisson's ratio $\nu$ & 0.3\\
	Total iterations & 200\\
	Projection $\beta$ & $\{2,4,8\}$ at iter. $\{0,50,100\}$\\
	Joint resultant stiffness $k_c$ & 10\\
	\noalign{\smallskip}\hline
\end{tabular}

\subsection{Parameters for 3D examples}\label{ap:3Dparams}

\begin{tabular}{lr}
	\hline\noalign{\smallskip}
	Parameter & Value\\
	\noalign{\smallskip}\hline\noalign{\smallskip}
	Element size & $1 \times 1 \times 1$\\
	SIMP penalty exponent $p$ & 3\\
	Filter radius $r$ & 1.5\\
	Young's modulus bracket & 1\\
	Young's modulus plate & 3\\
	Poisson's ratio $\nu$ & 0.3\\
	Total iterations & 200\\
	Projection $\beta$ & $\{2,4,8\}$ at iter. $\{0,50,100\}$\\
	Joint resultant stiffness $k_c$ & 10\\
	\noalign{\smallskip}\hline
\end{tabular}

\bibliography{EBHF}%

\begin{thebibliography}{10}
\providecommand \doibase [0]{http://dx.doi.org/}%

\bibitem{emmelmann_10_2017}
Emmelmann C, Herzog D, Kranz J. {Design} for laser additive manufacturing. In:
  Brandt M. \kern-2pt, ed. {\it Laser {Additive} {Manufacturing}}Woodhead
  {Publishing} {Series} in {Electronic} and {Optical} {Materials}. Woodhead
  Publishing.  2017 (pp. 259--279)

\bibitem{bruns_topology_2001}
Bruns TE, Tortorelli DA. Topology optimization of non-linear elastic structures
  and compliant mechanisms. {\it Computer Methods in Applied Mechanics and
  Engineering} 2001\string; 190(26)\string: 3443--3459.
\newblock \href {\doibase 10.1016/S0045-7825(00)00278-4} {doi:
  10.1016/S0045-7825(00)00278-4}

\bibitem{bendsoe_topology_2004}
Bends{\o}e MP, Sigmund O. {\it Topology {Optimization} {Theory}, {Methods}, and
  {Applications}}.
\newblock Springer-Verlag Berlin Heidelberg.
\newblock 2~ed. 2004.

\bibitem{ambrozkiewicz_density-based_2020}
Ambrozkiewicz O, Kriegesmann B. Density-based shape optimization for fail-safe
  design. {\it Journal of Computational Design and Engineering} 2020\string:
  accepted for publication.

\bibitem{huang_evolutionary_2010}
Huang X, Xie M. {\it Evolutionary {Topology} {Optimization} of {Continuum}
  {Structures}: {Methods} and {Applications}}.
\newblock John Wiley \& Sons .
\newblock 2010.
\newblock Google-Books-ID: oNqwqj2u3FoC.

\bibitem{kreisselmeier_application_1983}
Kreisselmeier G, Steinhauser R. Application of vector performance optimization
  to a robust control loop design for a fighter aircraft. {\it International
  Journal of Control} 1983\string; 37(2)\string: 251--284.
\newblock \href {\doibase 10.1080/00207179.1983.9753066} {doi:
  10.1080/00207179.1983.9753066}

\bibitem{bourdin_filters_2001}
Bourdin B. Filters in topology optimization. {\it International Journal for
  Numerical Methods in Engineering} 2001\string; 50(9)\string: 2143--2158.
\newblock \href {\doibase 10.1002/nme.116} {doi: 10.1002/nme.116}

\bibitem{zhu_integrated_2009}
Zhu J, Zhang W, Beckers P. Integrated layout design of multi-component system.
  {\it International Journal for Numerical Methods in Engineering} 2009\string;
  78(6)\string: 631--651.
\newblock \href {\doibase 10.1002/nme.2499} {doi: 10.1002/nme.2499}

\bibitem{sigmund_topology_2013}
Sigmund O, Maute K. Topology optimization approaches. {\it Structural and
  Multidisciplinary Optimization} 2013\string; 48(6)\string: 1031--1055.
\newblock \href {\doibase 10.1007/s00158-013-0978-6} {doi:
  10.1007/s00158-013-0978-6}

\bibitem{thomas_topology_2020}
Thomas S, Li Q, Steven G. Topology optimization for periodic multi-component
  structures with stiffness and frequency criteria. {\it Structural and
  Multidisciplinary Optimization} 2020.
\newblock \href {\doibase 10.1007/s00158-019-02481-7} {doi:
  10.1007/s00158-019-02481-7}

\bibitem{bendsoe_optimal_1989}
Bends{\o}e MP. Optimal shape design as a material distribution problem. {\it
  Structural optimization} 1989\string; 1(4)\string: 193--202.
\newblock \href {\doibase 10.1007/BF01650949} {doi: 10.1007/BF01650949}

\bibitem{chickermane_design_1997}
Chickermane H, Gea HC. Design of multi-component structural systems for optimal
  layout topology and joint locations. {\it Engineering with Computers}
  1997\string; 13(4)\string: 235--243.
\newblock \href {\doibase 10.1007/BF01200050} {doi: 10.1007/BF01200050}

\bibitem{zhu_multi-point_2015}
Zhu JH, Gao HH, Zhang WH, Zhou Y. A {Multi}-point constraints based integrated
  layout and topology optimization design of multi-component systems. {\it
  Structural and Multidisciplinary Optimization} 2015\string; 51(2)\string:
  397--407.
\newblock \href {\doibase 10.1007/s00158-014-1134-7} {doi:
  10.1007/s00158-014-1134-7}

\bibitem{jansen_topology_2013}
Jansen M, Lombaert G, Schevenels M, Sigmund O. Topology optimization of
  fail-safe structures using a simplified local damage model. {\it Structural
  and Multidisciplinary Optimization} 2013\string; 49(4)\string: 657--666.
\newblock \href {\doibase 10.1007/s00158-013-1001-y} {doi:
  10.1007/s00158-013-1001-y}

\bibitem{rakotondrainibe_topology_2020}
Rakotondrainibe L, Allaire G, Orval P. Topology optimization of connections in
  mechanical systems. {\it Structural and Multidisciplinary Optimization} 2020.
\newblock \href {\doibase 10.1007/s00158-020-02511-9} {doi:
  10.1007/s00158-020-02511-9}

\bibitem{svanberg_method_1987}
Svanberg K. The method of moving asymptotes - a new method for structural
  optimization. {\it International Journal for Numerical Methods in
  Engineering} 1987\string; 24(2)\string: 359--373.
\newblock \href {\doibase 10.1002/nme.1620240207} {doi: 10.1002/nme.1620240207}

\bibitem{wang_projection_2011}
Wang F, Lazarov BS, Sigmund O. On projection methods, convergence and robust
  formulations in topology optimization. {\it Structural and Multidisciplinary
  Optimization} 2011\string; 43(6)\string: 767--784.
\newblock \href {\doibase 10.1007/s00158-010-0602-y} {doi:
  10.1007/s00158-010-0602-y}

\end{thebibliography}

\end{document}